# Sustainable steel through hydrogen plasma reduction of iron ore: process, kinetics, microstructure, chemistry

I. R. Souza Filho[1,*], Y. Ma[1], M. Kulse[1], D. Ponge[1], B. Gault[1,2], H. Springer[1,3], D. Raabe[1]

[1] *Max-Planck-Institut für Eisenforschung, Max-Planck-Str. 1, 40237 Düsseldorf, Germany*
[2] *Department of Materials, Imperial College London, South Kensington, London SW7 2AZ, UK*
[3] *Institut für Bildsame Formgebung, RWTH Aachen University, Intzestr. 10, 52072 Aachen, Germany*

*corresponding author: i.souza@mpie.de

**Abstract**

Iron- and steelmaking is the largest single industrial $CO_2$ emitter, accounting for 6.5% of all $CO_2$ emissions on the planet. This fact challenges the current technologies to achieve carbon-lean steel production and to align with the requirement of a drastic reduction of 80% in all $CO_2$ emissions by around 2050. Thus, alternative reduction technologies have to be implemented for extracting iron from its ores. The hydrogen-based direct reduction has been explored as a sustainable route to mitigate $CO_2$ emissions, where the reduction kinetics of the intermediate oxide product $Fe_xO$ (wüstite) into iron is the rate-limiting step of the process. The total reaction has an endothermic net energy balance. Reduction based on a hydrogen plasma may offer an attractive alternative. Here, we present a study about the reduction of hematite using hydrogen plasma. The evolution of both, chemical composition and phase transformations was investigated in several intermediate states. We found that hematite reduction kinetics depends on the balance between the initial input mass and the arc power. For an optimized input mass-arc power ratio, complete reduction was obtained within 15 min of exposure to the hydrogen plasma. In such a process, the wüstite reduction is also the rate-limiting step towards complete reduction. Nonetheless, the reduction reaction is exothermic, and its rates are comparable with those found in hydrogen-based direct reduction. Micro- and nanoscale chemical and microstructure analysis revealed that the gangue elements partition to the remaining oxide regions, probed by energy dispersive spectroscopy (EDS) and atom probe tomography (APT). Si-enrichment was observed in the interdendritic fayalite domains, at the wüstite/iron hetero-interfaces and in the primarily solidified oxide particles inside the iron. With proceeding reduction, however, such elements are gradually removed from the samples so that the final iron product is nearly free of gangue-related impurities. Our findings provide microstructural and atomic-scale insights into the composition and phase transformations occurring during iron ore reduction by hydrogen plasma, propelling better understanding of the underlying thermodynamics and kinetic barriers of this essential process.

# 1. Introduction

Annually, 2.6 billion tons of iron ore (mostly hematite) are converted into steel by the integrated blast furnace (BF) and basic oxygen furnace (BOF) route [1], accounting for approximately 70% of the global steel production. The remaining 30% is realized by melting steel scraps and directly reduced iron (the latter is also referred to as iron sponge) in electric arc furnaces (EAF). On average, about 2.1 tons of $CO_2$ are produced per ton of crude steel [2]. This number corresponds to about 6.5% of all $CO_2$ emissions on the planet and makes iron- and steelmaking the largest industrial individual emitter of $CO_2$ [3]. To mitigate global warming, a drastic reduction of 80% in all $CO_2$ emissions is targeted until 2050 [4–6]. Thus, disruptive technology changes in iron ore reduction must be urgently implemented, already within the next years.

The use of hydrogen instead of carbon for iron ore reduction is currently explored as an alternative sustainable route to mitigate the $CO_2$ emissions [7–9]. The direct reduction of iron ore pellets by pure molecular hydrogen above 570°C occurs with the intermediate formation of other iron oxide variants, viz. $Fe_2O_3$ (hematite) → $Fe_3O_4$ (magnetite) → $Fe_xO$ (wüstite) → Fe (iron) [7–22]. Although the net energy balance of the overall process is endothermic [10], studies demonstrate that its reduction kinetics can be reasonably faster than that of commercial direct reduction conducted in shaft furnaces using reformed natural gas (e.g. the Midrex process) [8,18]. This observation motivates the current global investments in hydrogen-based direct reduction pilot plants [9].

Hydrogen plasma offers a viable alternative for carbon-neutral iron making. Its high energy and enhanced density of H radicals and exited states help to overcome the reaction's activation barrier [23–27] and has the potential for enhancing the $Fe_xO$ reduction rates by an order of magnitude [28–30], enabling iron conversion to reach commercially viable rates. Also, hydrogen plasma-based reduction allows the production of liquid iron in one single step, in which the input fine ores are melted and reduced simultaneously without the need for intermediate agglomeration or refinement processing, as the melting point of iron oxide (1565°C) only slightly exceeds that of iron (1538°C) [31,32].

During the hydrogen plasma reduction (HPR), a plasma arc zone is generated between an electrode and the input iron ore (e.g. hematite) [33]. In this zone, the ore can be melted and reduced by hydrogen in both molecular and plasma states. The latter is composed of vibrationally ionized ($H^+$, $H_2^+$ and $H_3^+$), excited ($H^*$) and atomized (H) species, which are formed through the mutual elastic and inelastic collisions of hydrogen particles with electrons [34,35]. Under such conditions, the corresponding degree of hydrogen dissociation is determined by the competing ionization (e.g. $e^- + H_2 \rightarrow H_2^+ + 2e^-$) and recombination (e.g. $H^+ + e^- \rightarrow H$) events [34]. The high energy carried by the hydrogen plasma is partially released at the reaction interface, generating a large amount of local heat [36,37]. This heat diminishes the need for external power supply and promotes high power-efficiency to the process [32]. Thus, the thermo-kinetics advantages observed in HPR depend directly on the concentration of hydrogen plasma radicals that are able to reach the reaction interface [25,34,36,37].

Plasma discharges are generated by electric arcs in conventional EAFs to melt steel scraps. Thus, it is conceivable that established industrial electric furnaces can be modified to be used for HPR purposes, with only small (e.g. 5-10%) $H_2$ partial pressure. Due to the limited number of scientific investigations on this topic [25,28,29,34–39], closer analysis of the underlying reaction steps and of the influence of gangue-related tramp elements at the nano-scale may help to identify conditions for developing HPR processes with high kinetics and efficiency.

Here, we performed interrupted Ar-10%$H_2$-plasma reduction experiments of hematite to study and evaluate the HPR process stepwise in high detail. Emphasis is placed on the evolving nano-chemistry, interface structure and composition and the phase transformation kinetics. We observe that the hematite reduction kinetics depends on the balance between input material mass and arc power. Wüstite reduction is the rate-limiting step of the process and the gangue elements are gradually removed from the samples with an increase in the reduction cycles. This seems to be particularly important for the case of Si, which in conventional H-based direct reduction can decorate and impede the hetero-interface between Fe and the $Fe_xO$. We show that low-impurity-content iron can be produced by hydrogen plasma with reduction rates comparable to those found in the direct reduction conducted using molecular hydrogen. Our findings advance the understanding of the succession of phase transformations and chemical evolution during HPR and their influence on the thermodynamic and kinetic barriers to scale up green iron production.

## 2. Experimental procedures
### 2.1 Input material and hydrogen plasma-based reduction

Commercial hematite pieces were used in this study and their chemical composition is presented in Table 1. The density of the hematite pieces was measured as 5.2983 ± 0.0016 g/cm$^3$ via argon pycnometer analysis. Samples of hematite with an average weight of 9 g were introduced into an arc-melting furnace, which was equipped with a tungsten electrode (diameter of 6.34 mm) and filled with a gas mixture of Ar-10%$H_2$ (total pressure of 900 mbar, chamber volume of 18 L). To melt and reduce the samples, a plasma arc was ignited between the tip of the electrode and the input material at a voltage of 44 V and a current of 800 A. The average distance between the electrode and the hematite samples was 6 mm. Fig. 1 (a) shows a schematic representation of the apparatus.

To track the progress of both, the phase transformations and chemical composition over the course of the reduction process, the samples were submitted to interrupted reaction cycles. To obtain specimens in different partially reduced states, the hematite pieces were subjected to sequential melting, solidification and remelting series at a pre-defined number of times (viz. 1, 2, 5, 10, 15 and 30 times) with a corresponding exposure time to the hydrogen plasma at 1 min for each cycle. During the 1-min cycles, complete melting of the samples was achieved with an average exposure time of 6 seconds. Once the plasma arc was switched off, rapid solidification conditions were achieved by means of the water-cooled copper hearth on which the hematite pieces were molten and reduced. To ensure the proper stoichiometry conditions (i.e. a reductant atmosphere) and maintain the progressing reduction, the furnace chamber was replenished by a fresh mixture of Ar-10%$H_2$ after the completion of every individual melting and solidification cycle. In summary, one single process cycle comprises the following steps: 1) charging of the furnace with the oxide sample and flooding of the chamber with fresh Ar-10%$H_2$ gas mixture; 2) simultaneous plasma arc melting and reduction of the sample; 3) intermediate interruption of the plasma arc operation and solidification of the (partially or fully reduced) sample, and 4) replenishing of the furnace's chamber with fresh Ar-10%$H_2$ gas mixture for the continuation of the process (i.e. succeeding cycle). The sequential steps of this procedure are summarized in Fig. 1 (b). Results will be reported as a function of the total time of exposure to the plasma experienced by each sample. The reduced samples (Fig. 1c) were further characterized via microstructural, chemical and phase transformation analysis, in part down to atomic scale.

**2.2 Microstructural characterization**

The solidified samples were cut, embedded and metallographically prepared for high-resolution scanning electron microscopy analysis, including electron backscatter diffraction (EBSD) and electron channeling contrast imaging (ECCI). EBSD maps were acquired using a JEOL-6500F field-emission gun microscope. The mapped areas were further imaged with the aid of the ECCI technique using a Zeiss Merlin scanning electron microscope (SEM). Energy-dispersive X-ray spectroscopy (EDS) analyzes were also conducted in representative areas of the samples, using a Zeiss Merlin SEM and an accelerating voltage of 15 kV.

**2.4 Chemical composition analysis**

To obtain the average weight percentage of elements at each stage of reduction, chemical composition analysis was conducted for the solidified samples containing both, the iron and untransformed oxide portions (Fig. 1c). The content of metals was determined using inductively coupled plasma optical emission spectrometry (ICP-OES). The oxygen content was determined via reduction fusion under a helium atmosphere and subsequent infrared absorption spectroscopy. Carbon and sulfur were analyzed via combustion followed by infrared absorption spectroscopy. Iron nuggets extracted from the partially reduced samples (5, 10, 15 min) were also analyzed.

**2.5 Thermodynamic simulations**

Based on the average chemical composition of representative samples, the corresponding metastability phase calculations, in which the face-centered cubic iron was excluded, were conducted using the software ThermoCalc, coupled with the metal oxide solutions TCOX10 database. For comparison, the corresponding phase equilibrium diagrams were also calculated and reported in the Supplementary Material (Fig. S.1).

**2.6 Atom probe tomography**

To characterize the atomic-scale elemental distribution across the wüstite-iron interface, atom probe tomography (APT) was employed. The APT specimens were prepared using the site-specific lift-out method at the wüstite-iron interface (Fig. S.2) of the 5 min-reduced sample by an FEI Helios NanoLab600i dual-beam focused ion beam/scanning electron microscopy instrument. The APT measurements were conducted using a Local Electrode Atom Probe (LEAP, Cameca Instruments Inc.) 5000 XS with a detection efficiency of ~ 80%. The instrument was operated in laser-pulsing mode at a wavelength of 355 nm, laser energy of 40 pJ, and laser pulse frequency of 200 kHz. The base temperature in the analysis chamber was maintained at 50 K during the measurements. The reconstruction of the three-dimensional atom maps and data analysis was carried out using the commercial software AP Suite 6.0.

**2.7 Phase quantification analysis**

The partially reduced samples were hammered to separate the formed iron portion, i.e. nuggets, from the remaining oxides. The obtained iron and oxide portions were weighted in a precision analytical balance (accuracy of ± 0.0001 g). The remaining oxide was further ground and probed by X-ray diffraction measurements (XRD) using a diffractometer D8 Advance A25-X1, equipped with a cobalt $K_\alpha$ X-ray source, operated at 35 kV, 40 mA. The diffraction data were processed using the software Brucker Topas v. 5.0. Rietveld refinement was employed to quantify the weight fraction of the constituents in the powder. To determine the iron conversion kinetics, the weight of the iron within the oxide portion was summed with that of the iron nugget extracted from the same sample.

**3. Results**

**3.1 Macrostructural analysis**

Figure 1 (c) shows the cross-section of the samples reduced with different exposure times to the hydrogen plasma. After 1 and 2 min of reduction, no visual evidence for iron is observed. From 5 to 15 min, millimeter-sized iron portions (indicated by the red arrows in Fig. 1c) appear brighter than the remaining oxides, the latter mostly located in the top part of the samples. Total conversion is reached after 30 min. Also, the average porosity of approximately 35% is found in the partially reduced samples (from 1 to 15 min). Details regarding the porosity analysis are given in the Supplementary Material (Fig. S.3). The completely reduced sample is free of macro-pores.

**3.2 Microstructural analysis**

Figure 2 shows the microstructural analysis of the specimen partially reduced for 1 min. The analyzed area is indicated in Fig. 2 (a) by the red frame. Fig. 2 (b) depicts the corresponding ECCI-image and reveals a columnar grain structure with hundreds of μm in length. The EBSD inverse pole figure (IPF) map in Fig. 2 (c) shows that the grains grow with a strong crystallographic texture with the majority of the <001> directions aligned perpendicular to the sample's surface (i.e. parallel to the solidification direction). Such a columnar microstructure is consistent with that observed also in other unidirectional solidification processes, governed by fast heat extraction, imposed here by the water-cooled hearth. The phase map displayed in Fig. 2 (d) shows the coexistence of magnetite and wüstite in this sample, represented in purple and green, respectively. The remaining magnetite, when found dispersed within wüstite, shows a cuboidal morphology, Figs. 2 (e-g).

Figure 3 (a) is a similar overview image of the sample partially reduced for 2 min. The analyzed area is indicated by the red frame in this figure. Fig. 3 (b) shows that small iron volumes with sizes of up to 10 μm are readily distinguished due to their bright contrast (arrowed). Fig. 3 (c) shows details of a drop-shape iron domain surrounded by smaller satellite iron islands. The same area was further probed via EBSD and the corresponding maps are displayed in Figs. 3 (d-h). Both, the EBSD phase (Fig. 3f) and oxygen distribution (Fig. 3g) maps confirm the presence of iron in the microstructure. Yet, they evidence that small oxide particles are found within iron (Fig. 3 f and g) as well as the wüstite/ferrite phase interfaces are enriched in silicon (Fig. 3h). For all investigated areas of this sample, wüstite was the remaining iron oxide phase.

Figure 4 shows the microstructural analysis of the sample partially reduced for 5 min. EBSD maps, shown in Fig. 4 (b), were acquired for the region indicated by the red frame in Fig. 4 (a). The oxygen distribution map in Fig. 4 (c) evidences small iron domains (arrowed) dispersed within wüstite. Also, Fig. 4 (c) reveals a Si enrichment at the wüstite/iron phase interfaces as well as in the interdendritic regions (backscatter electron image, BSE). The small volume fraction of this constituent hindered its quantification via XRD. However, similar analysis conducted for the sample partially reduced for 10 min (Supplementary Material, Fig. S.6) suggests that it is fayalite ($Fe_2SiO_4$).

Oxide particles are observed within iron, as revealed by the secondary electron (SE) image in Fig. 4 (d). Their chemical composition can be qualitatively evaluated via the corresponding EDS maps. Fig. 4 (d) shows a large particle (> 1 μm) that has a Si-enriched core surrounded by a Fe-O-bearing shell. The smaller particles are composed of iron and oxygen. The formation mechanisms of these oxide particles can be associated with the reduction kinetics and with hydrodynamic aspects of the molten bath flow. In terms of kinetics, one can assume that the molten oxide, when in direct contact with the plasma, does not in some cases get entirely reduced. In this scenario, small oxygen-enriched molten pools remain partially reduced within the volume majorly composed of liquid iron. When considering that the interaction between the arc plasma and the molten material is a rather complex phenomenon, one might also expect that turbulences of the melt flow would allow partially reduced domains (e.g. Fe$_x$O) to get trapped inside the liquid iron, due to bath dynamics. The fast solidification rate imposed by the water-cooled copper hearth might be a reason that these molten oxygen-enriched (i.e. partially reduced) pools solidify in the form of oxide particles. In this case, their chemistry should not vary drastically from that of the larger oxide portions located in the upper part of the sample (composed of wüstite and fayalite). Thus, we could plausibly expect compositions ranging from Fe$_x$O to Fe$_2$SiO$_4$ for these particles as well.

Figure 5 shows the atom probe tomography analysis of a wüstite-iron interface in the sample partially reduced for 5 min. The cumulative field evaporation histogram is displayed in Fig. 5 (a), and the red arrows highlight the wüstite-iron interface. The reduced iron reveals the crystallographic poles in the field evaporation histogram. The crystallographic pole marked by the red dot is identified as (011). The corresponding {011} atomic planes are visualized in Fig. 5 (b) and the interplanar spacing is approximately 0.20 nm. This value matches well with the interplane spacing of body-centered cubic iron. Such crystallographic information is also helpful for the improvement of the spatial accuracy of the reconstruction of the 3D atom maps [40,41]. Fig. 5 (c) shows the distribution of iron and oxygen atoms in the APT specimen. The interface of the wüstite and iron is marked by a 20 at.% oxygen iso-composition surface. The reduced iron contains more than 98 at.% iron, as shown by the composition profile calculated across the interface in Fig. 5 (d) and summarized in Table 2. In contrast, the APT analysis reveals that wüstite is mainly composed of ~ 59 at.% iron and 40 at.% oxygen, yielding a stoichiometric ratio ~ 3:2, which differs from the theoretically expected stoichiometric ratio ~ 1:1. The lower oxygen concentration measured by APT is most likely due to the loss of neutral oxygen during field evaporation as desorbed neutral oxygen atoms and also O$_2$ molecules, which cannot be measured by the detector [42–44]. The contents of trace elements (e.g. Al, Ca, Mg, Si, Co, Ni, and V) are considerably lower in the reduced iron compared to those within the oxide, as shown in Table 2. In addition, a sharp enrichment of silicon is observed at the wüstite/bcc-iron interface, as demonstrated by the concentration profile across the interface in Fig. 5 (e).

Figure 6 shows the microstructural evaluation of the sample completely reduced into iron (30 min). Fig. 6 (b) shows an ECCI-image which reveals columnar grains subdivided by sub-grains. The microstructure is nearly devoid of oxide particles. The enlarged view provided by Fig. 6 (c) reveals a few worm-like inclusions, whose volume percentage is only 0.10 ± 0.05 %. A representative SE-EDS analysis for one of these inclusions (Figs. 6 d-g) suggests that they are enriched in Si and O. The point EDS analysis performed within it (yellow star in Fig. 6h) also reveals the presence of several gangue elements, including Ca, Si, S and Cu, Fig. 6 (i) [45].

### 3.2.1 Oxide particles within iron

The evolution of both, the size and fraction of oxide particles within iron accompanies the reduction progress, as demonstrated by Fig. 7 (a). This figure shows that the particle percentage reaches up to 3.3 ± 0.6 % with 15 min of reduction, but sharply drops to nearly zero after 30 min (complete reduction). The weighted average size of the particles evolves up to 0.95 μm with 15 min, but it is considerably reduced to 0.16 μm in the final iron product. Figs. 7 (b) and (c) compare the microstructures of the iron portions obtained with 15 and 30 min of reduction, respectively. The final stages of the process (i.e. from 15 min onwards) seem to enable the deoxidation of iron.

### 3.3 Chemical analysis

Figure 8 shows the changes in the chemical composition of the samples during reduction. Here, it is important to clarify that these values represent the chemical composition of the entire specimen without distinguishing the iron and untransformed oxide portions. For the sake of comparison, the chemical composition of the initial hematite is also displayed in this figure (viz. at 0 min). Fig. 8 (a) shows that the oxygen content of the samples gradually decreases until reaching 0.0064 wt.% (64 ppm) in the final iron product.

An interesting general trend is that the quantities of Al, Mn and Si (Fig. 8 b), P and S (Fig. 8 d) are remarkably reduced over the course of the reduction. This purification process can be achieved by sputtering induced by the plasma arc and/or by removal via gas phases [32]. Although C, Ca and Mg concentrations slightly increase up to 15 min of reduction, the final stages of the reduction drop the corresponding Ca and Mg values to 10 ppm (Fig. 8 c). The C content in the final iron was found to be slightly higher than that in the initial hematite.

Cu and W enrichment is observed up to 2 min of reduction as a likely consequence of the melting of small parts of the hearth and electrode (Fig. 1a), respectively (Fig. 8e). For longer times, the Cu content is reduced to 20 ppm, which is lower than that found in hematite (50 ppm). W values in turn vary with a non-defined trend.

The chemical composition evolution of the iron portion formed from 5 min of reduction onwards is given in Table 3. This result shows that the produced iron is nearly free of harmful elements such as P and S (< 20 ppm). Yet, the gradual deoxidation is essential to the removal of oxide particles, as previously observed in Fig. 7. The gangue elements are residual and, according to Fig. 6, they are confined in the negligible fraction of the worm-like inclusions.

### 3.4 Thermodynamic simulations

Based on the chemical composition of the samples partially reduced for 5, 10 and 15 min (Fig. 8), the corresponding metastability of the phases were calculated as displayed in Figs. 9 (a-b), respectively. These diagrams enable us to infer the likely metastable solidification path of the samples. Due to the low contents of P and S (Fig. 8), these elements were not considered in the calculations. W and Cu are incorporated as tramp elements during the melting and thus were also neglected.

Figure 9 shows that, under all reduction conditions probed in this study, the Fe-enriched liquid starts solidifying into bcc-Fe (body-centered cubic iron) at approximately 1530ºC, whereas the coexistent oxygen-enriched liquid portion is replaced by wüstite at temperatures below 1370ºC. At the final stages of the solidification, the remaining liquid is transformed into fayalite, as indeed observed in the interdendritic regions inside wüstite (Fig. 4 and Fig. S.4). Figs. 9 (b) and (c) also show that fayalite can decompose, via a solid-state transformation, into two variants (viz., one enriched in iron and the other doped with Ca) at temperatures around 1000°C. The weight fractions of bcc-Fe and wüstite displayed in Fig. 9 are in excellent agreement with the ones experimentally determined and reported in Fig. 8.

For comparison, the corresponding equilibrium phase diagrams are shown in Fig. S.1 of the Supplementary Material. Fig. S.1 shows that the solid-state bcc-iron → fcc-iron transformation is expected to occur between 910 and 1390°C. However, due to the very fast solidification conditions imposed by the water-cooled copper hearth, such transformation is kinetically unfavored to proceed in practice.

### 3.5 Phase transformations

The phase transformations as a function of the plasma exposure time are shown in Fig. 10. In this figure, only the major Fe/O-containing phases are considered, i.e. the very small fraction of the interdendritic fayalite (0.73 wt.%, Supplementary Material) within $Fe_xO$ was neglected. Fig. 10 (a) shows that hematite is reduced into a dual-phase magnetite (0.24) and wüstite (0.76) mixture after 1 min of reduction, confirming the previous observations made via microstructural analysis (Fig. 2). With 2 min of reduction, most of the sample is wüstitic (0.98) and the metallic iron drops observed for this reduction condition (Fig. 3) represent a weight fraction of 0.02. For longer reduction times, wüstite is gradually transformed into iron, whose formation kinetics follows a sigmoidal trend, Fig. 10 (b). The complete reduction is achieved after 30 min and the corresponding iron conversion rate reaches a maximum (8 wt.%/min) at approximately 3.5 min.

## 4. Discussion
### 4.1 Iron conversion kinetics

Figure 10 (a) showed that the last 15 min of reduction via hydrogen plasma are accompanied by relative sluggish kinetics. In other words, 75% reduction is achieved within the initial period of 15 min, whereas the last 25% also requires 15 min. This observation indicates that the intermediate wüstite reduction imposes thermodynamic and kinetic limiting barriers to the total conversion into iron. In fact, the reduction of molten wüstite by hydrogen plasma is the most unfavorable reaction when compared with the preceding reduction steps, at temperatures ranging from 1600 to 2000°C [34,36]. For example, the equilibrium Gibbs free energy changes ($\Delta G$) for reducing wüstite, magnetite and hematite using $H^+$ ions at 1600°C are approximately $-2.9 \cdot 10^{-3}$ kJ/mol, $-3.9 \cdot 10^{-3}$ kJ/mol and $-4.4 \cdot 10^{-3}$ kJ/mol, respectively [34]. These energy values are generally calculated assuming that all reductant species in the arc plasma are $H^+$ ions in thermodynamic equilibrium, i.e. neglecting all the plasma particle collisions. In contrast to this idealized exothermic plasma scenario, one must bear in mind that the solid state wüstite reduction using molecular hydrogen (i.e. direct reduction) displays positive $\Delta G$ values and, therefore, this net process is thermodynamically not exothermic [34,36,46]. This means that additional energy must be provided to drive the underlying redox reaction. This endothermic energy balance of the oxide reduction in the solid / gas mixture advocates the use of a hydrogen plasma instead of regular molecular hydrogen, as the energy provided to create a plasma produces a high density of very reactive radicals that overcome the thermodynamic barriers readily and equips the overall wüstite reduction with high efficiency.

Figure 11 shows the diagram 'microstructure versus oxygen content' for the reduced samples. This diagram helps to observe how the total oxygen content in a sample is partitioned into different phases. The very small oxygen fraction found in both, the residual fayalite and oxide particles within iron were not considered in the results displayed in this figure. Based on the data shown in Fig. 11, we fairly estimated the oxygen removal after each reduction stage and the obtained results are displayed in Table 4. The amount of hydrogen not consumed during the reaction was also estimated through the calculation of mass balance, taking into account the oxygen content in the samples (neglecting the small material loss induced by arc spattering), their respective phase fractions and the hydrogen provided in the furnace chamber after each cycle, as also depicted in Table 4.

It was proposed [28] that reduction reactions in HPR occur through the following steps: (1) mass transfer of hydrogen and oxygen respectively from the gas and molten oxide to the reaction interface; (2) adsorption of hydrogen and adsorption/dissociation of $Fe_xO$ at the reaction interface; (3) chemical reaction; (4) desorption of $H_2O$ from the interface; (5) mass transfer of $H_2O$ away from the interface. Ref. [28] also reveals that there is a critical value of gas flow rate above which the mechanism of gas transfer is not the rate-limiting step of the process (i.e. the reduction rate becomes independent on the gas flow rate). If the furnace's atmosphere contains a stoichiometrically sufficient amount of reductant agents, the overall reduction rate is determined by the slowest step which is assumed to be step (3), i.e. the chemical reaction. Differently from Ref. [28], our experiments were not conducted under constant gas flux. However, by replenishing the furnace chamber with fresh Ar-10%$H_2$ gas mixture after the completion of one melting cycle (Section 2.1 and Fig. 1b), we ensure sufficient stoichiometry conditions for the progressing reaction, since the provided amount of hydrogen is much higher than that consumed during each reaction step (Table 4). Yet, according to the Baur-Glaessner diagram modified to consider the atomic hydrogen existing at high temperatures in the plasma medium, iron is stable in the presence of at least 60% of $H_2$ molecules contained in a mixture of $H_2O + H_2$ (viz. product water and excess of hydrogen), between 1600 and 2300°C [32]. The corresponding '$H_2$ in ($H_2O + H_2$)' molar fractions at each reduction time are also given in Table 4 and evidence that the furnace atmosphere was always kept reducing, thus displaying favorable conditions for iron formation. Therefore, appropriate stoichiometry is provided, and the plasma generation conditions are identical (44 V, 800 A). Due to these well-controlled chemical boundary conditions, we can plausibly infer that the mass transfer from the gas to the reaction interface has been similar during all melting cycles, providing similar conditions to the reaction kinetics [28,29].

The relative sluggish kinetics of wüstite reduction from 15 min onwards is associated with its ever-lower fractions, i.e. the lower the content of wüstite in the sample the more difficult is its reduction. This fact is most probably linked to physical characteristics of the plasma arc process: the reduced iron has higher mass density than the remaining wüstite. Consequently, iron tends to sink to the bottom of the melt, leaving remaining wüstite on top (similar to a slag) which gets thus exposed to the arc plasma (as reduction can only occur on the melt surface). However, the density difference is not so significant that it can sufficiently counteract the dynamic material flow occurring in the melt. Such a dynamic flow is caused by the forces (magnetic and kinetic) imposed by the arc, together with the localized heating on top (induced by the contact point of the arc) and cooling on bottom of the melt (by the water-cooled hearth). While this fact is favorable from a point of desired homogeneity of the melt, it also constantly moves the unreduced wüstite through the desired reaction zone directly under the plasma arc, which, with decreasing wüstite fraction, becomes less and less probable. Furthermore, as the wüstite fraction and oxide particle size decreases, the arc plasma preferably contacts already reduced Fe, which exhibits less resistance to the electron transfer required [34,35,38,39]. These factors strongly depend on the respective experimental conditions and can be addressed by technical improvements (especially in an industrial EAF, which typically uses lower temperature arc plasmas and lower arc power to melt volumes) such as pulsed arcs, external magnetic fields or deliberate pauses between arc cycles to induce oxide floatation. In light of these observations, the most probable rate-limiting barriers after 75% of reduction (Fig. 10 a) can be considered to be the oxygen mass transportation to the reaction interface and the reduction reaction itself.

The reduction kinetics of molten oxides are dependent on several experimental factors such as the electrode's material and diameter, length of the arc, ratio of the input mass to the arc power and cooling. To provide better conditions for wüstite floatation and ensure its exposure to the reaction zone, we increased the input material weight to approximately 15 g, while all the other parameters were kept constant. This second reduction procedure was conducted as described in Section 2.1 and the obtained phase evolution is displayed in Fig. 12 (a). This figure shows that already 1 min of plasma exposure promotes total transformation of hematite into wüstite. For longer times, the iron conversion follows a sigmoidal trend, and total reduction is achieved after 15 min.

The maximum transformation rate (12 wt.%/min) occurs at approximately 7 min (Fig. 12b). This result means that the reduction of the 15 g-hematite pieces proceeds two times faster than that of the 9 g pieces, yet, at rates of about one order of magnitude higher.

This second set of experiments confirms that a substantial kinetics enhancement can be achieved by adjusting the input material mass, the mass-to-surface ratio, and consequently the molten bath height, in relation to the plasma arc power and general furnace conditions (bath dimensions, cooling power, etc.). It appears that a larger molten bath facilitates mass transport of oxygen to the reaction interface (i.e. promoting wüstite flotation), thus speeding up the reaction. In this context, one must also consider that the hydrodynamic features developed during plasma arc melting are rather complex and several other parameters such as gas shear stress, Marangoni and electromagnetic forces, buoyancy, and thermo-capillarity convection contribute to mass and heat distribution throughout the liquid [31,38,47–52]. For the second set of reduced samples, 1 min of exposure to the hydrogen plasma enables the complete reduction of hematite into wüstite whereas additional 14 min is required to reach pure iron. This result confirms that the wüstite reduction is the rate-limiting step in the hydrogen plasma-based reduction of iron ores, even under conditions of optimized kinetics.

**4.2 Comparison with the kinetics data reported in the literature**

Seftejani et al. [35] investigated the reduction rates of iron ores in a hydrogen-plasma apparatus using a graphite electrode. At the early stages of their experiments, reduction was mainly achieved through carbon-containing agents stemming from the graphite electrode. Later, hydrogen acted as the major reductant. The reduction kinetics was determined via chemical analysis of the output gases (mass spectrometry) and the degree of metallization of the product reached values of approximately 70% (the remaining 30% was composed of slag). The reduction kinetics of iron ores using hydrogen plasma was also studied by Kamiya et al. [28]. Using a constant flow of an argon-hydrogen mixture (15% $H_2$) in a tungsten-equipped reactor, these authors found a reduction rate of 1.2 g/min. Our current experiments show that the iron conversion rates for the first set of experiments (using 9 g hematite pieces) varied from 8 %wt./min (at 3.5 min) to 1.2 %wt./min (at 30 min), Fig. 10 (b). Considering the input mass of 9 g, the corresponding rates of mass loss (0.72 and 1.08 g/min) are slightly lower than those reported in Ref. [28]. However, we had also found that the conversion rates were substantially enhanced when using hematite samples of higher mass (15 g). In this case, the maximum rate of 12 wt.%/min is translated into 1.8 g/min, a value which is 1.5 times higher than that found in Ref. [28]. Hence, our results show conversion rate values of a magnitude that has been reported also before for hydrogen plasma-based reduction of iron oxides. We also observe that faster reduction kinetics is achievable when the input material mass is properly adjusted in relation to the power supply provided by the plasma arc [53]. This finding suggests to devote more systematic studies to the influence of the ratio between ore mass, hydrogen partial pressure and plasma parameters in terms of voltage, power and current density.

## 4.3 Hydrogen direct solid-state reduction versus hydrogen plasma reduction of molten iron ore

The reduction kinetics of iron ores using different reductant agents ($H_2$, CO, solid carbon, Fe-C melt) was evaluated by Nagasaka et al. [29] and several other authors [54–62]. These authors reported that the reduction rates using hydrogen can be one or two orders of magnitude faster than those of redox reactions conducted with C-containing reductants. Recently, we investigated the direct (solid state) reduction of iron ore pellets under pure $H_2$ flow (0.5 L/min) at 700ºC, using a thermogravimetry experiment [21]. In that work, we chose a relative low temperature compared with those used in industrial shaft furnaces (viz. 850-1000°C) to monitor the intermediate microstructural changes prior to complete reduction and associate them with the rate-limiting steps of this process. Solid-state direct reduction of iron ore pellets using molecular hydrogen at higher temperatures has been reported in [15,18,55]. Turkdogan and Vinters [15] investigated the direct reduction of ~ 15 mm-pellets at 1000°C under several gas flow rates. Bonalde et al. [18] evaluated the reduction reaction at 850°C using different gases, including pure molecular hydrogen. Results reported in [15,18] thus provide insights into the reduction kinetics close to those targeted in the industrial practice. The reduction kinetics reported in these preceding works are reproduced as reference data in Fig. 13 (a) in comparison with the iron weight fractions obtained in this work via hydrogen plasma-based reduction. In this figure, a logarithmic abscissa is used to facilitate visualization and comparison of the data.

Figure 13 (a) reveals that a total reduction via the direct reduction process is achieved after 100 min of exposure to molecular hydrogen at 700ºC. From our preceding work, we found that the direct reduction occurs preferentially at the external layers of the pellet due to the short diffusion distances for O from the interface reaction to the surface.

Thus, at 33% reduction, iron is predominantly formed at the shell of the pellet whereas the core remains predominantly composed of wüstite. The further 67% of reduction is essentially given by the $Fe_xO + H_2 \rightarrow xFe + H_2O$ reaction and displays notably sluggish kinetics, as reproduced in Fig. 13 (b). This is particularly due to the slow nucleation of the iron crystals (thermodynamic barrier) and to the sluggish mass transport, particularly of the outbound transport of oxygen, throughout the compact iron layer that forms as a dense shell on the wüstite (i.e. as outer layers of the pellets) [21]. Thus, microstructure features such as pore tortuosity, formation of dislocations, internal interfaces and free surfaces due to volume mismatch among the phases, and nano-dispersed transient metal oxides were found to exert an important role in the overall reduction kinetics by providing additional free paths for reductant and product transport and fresh surfaces for reaction.

Figure 13 (a) also shows that complete reduction occurs after 15 min of direct exposure to hydrogen at 850°C ($H_2$ flow of 2 L/min) [18]. A similar trend is observed when the process is conducted at 1000°C with a $H_2$ flow rate of 1 L/min [15]. The reduction kinetics can be further enhanced at 1000°C by increasing the gas flow to 10 L/min so that the total conversion into iron is obtained within 10 min [15]. This behavior is most likely due to the faster diffusion of oxygen to the reaction interface at 850 and 1000°C rather than that at 700°C. It should be also noted that the magnitude of the $H_2$ flow rates used in Refs. [15,18] is at least twice as high as the one used in our preceding work [21] (0.5 L/min at 700°C, brown line in Fig. 13a), providing high mass transfer rates of reactants to the reaction interface. In addition, Turkdogan and Vinters [15] also observed that the pore structure of the outer iron layer formed at 1000°C is coarse and not so fine and highly dispersed as the ones observed in pellets reduced at lower temperatures (e.g. 700°C). The coarse pore structure of iron facilitates gas diffusion thorough it, thus helping to overcome the rate-limiting barrier of the process as well.

Figure 13 (b) shows the corresponding reduction rates for both, the hydrogen molecular-based solid state reduction and the hydrogen plasma-based reduction, where the oxide gets melted. The data reveal that the reduction rates through hydrogen plasma exposure are comparable with those observed through conventional hydrogen direct solid-state reduction conducted at temperatures typically employed in shaft furnaces (850-1000°C) and under moderate $H_2$ flow rates. It is worth to note though that the kinetic barriers associated with direct reduction (solid-state nucleation and solid-state mass transport) do not exist in plasma reduction, as in the latter process all reactions occur in the liquid state. Yet, hydrogen plasma species substantially reduce the values of ΔG for wüstite reduction (Section 4.1). The high reactivity of the radicals and the high energy content of the hydrogen plasma is transferred to the reaction interface (molten pool), enabling local heat release and self-supplying energy for the endothermic $Fe_xO$ reduction [36,37], which means that the wüstite reduction using hydrogen plasma is exothermic and not endothermic as the solid-state direct reduction. Thus, the exothermic net energy balance of the hydrogen plasma reduction ensures higher efficiency in power consumption when compared with that in the endothermic solid-state direct reduction, which requires additional external energy input to maintain the progressing reaction. All these factors explain why the reduction process via hydrogen plasma proceeds similarly to that of molecular hydrogen based direct reduction and reveal their advantages in terms of energy consumption.

**4.4 Micro- and nanoscale chemical evolution**

In the partially reduced samples, the gangue elements are mostly partitioned to the untransformed oxide when probed at the micro- and nanoscale, as revealed by the EDS and APT measurements, respectively. However, the chemical composition analysis (Fig. 8) reveals that such gangue elements are gradually eliminated from the samples over the course of the reduction, i.e. the plasma reduction process cleans the material, an advantage that is very important for instance when reducing high Si, P and S containing ores. The effect is mainly due to the relatively high vapor pressures of most of the gangue elements that occur in iron ores, relative to that of iron, revealing yet another thermodynamic advantage of the plasma over the conventional direct reduction process [31,32,63]. From Fig. 6 we can learn that the last 15 min are essential in the overall plasma-based reduction process for cleaning the formed iron from its retained oxide particles. The final iron product is nearly free of gangue residues, especially the harmful elements P and S have been removed (Figs. 6 and 8). These results confirm preceding observations that dephosphorization and deoxidation occur over plasma-reduction mainly through evaporation, thus reducing the need for expensive and time-consuming secondary metallurgy operations for removing these elements [32,64].

Silicon is found in the interdendritic fayalite, at the wüstite/α-iron phase interfaces and within the primarily solidified oxide particles pinned inside α-iron. Interdendritic fayalite forms in the final stages of solidification, as suggested by the corresponding thermodynamics simulations (Fig. 9) and as visualized in Section 3. Si-enrichment at the hetero-interfaces can also occur during solidification, when the solidification front of the wüstite encounters the solid iron. This finding can be attributed to the high diffusivity of silicon, which leads to its partitioning to the ever-decreasing fractions of liquid. Although silicon is quite abundant in the microstructure, its content also decreases during the course of reduction (Fig. 8). In this context, the reduction of the anionic complex $SiO_4^{-4}$ into metallic silicon is not promoted and it is eliminated from the samples through the gaseous SiO suboxide [32,64].

## 5 Summary and conclusions

We investigated the chemistry and phase evolution during hydrogen plasma-based reduction of commercial hematite. The following conclusions are drawn:

1) The iron conversion kinetics is dependent on the balance between the initial hematite mass and the arc power. The reduction kinetics of the 9 g hematite specimens is relative sluggish during the final 25% of reduction. By increasing the weight of the initial hematite pieces to 15 g, the reduction kinetics is substantially increased, occurring two times faster. The maximum conversion rate was 12 wt.%/min, which is translated into 1.8 g/min. Total reduction to pure iron was achieved after 15 min. It appears that larger molten baths facilitate transport of oxygen to the reaction interface. For both sets of experiments, using 9 g and 15 g-pieces, wüstite reduction is the rate limiting thermodynamic barrier of the process.

2) The reduction rate of hydrogen plasma-based reduction, using only a small $H_2$ partial pressure of 10%, is comparable with those ones observed in solid-state direct reduction of hematite conducted at temperatures currently employed in shaft furnaces (850-1000°C). The rate-limiting barriers (i.e. diffusion of oxygen species in the solid state) in the direct reduction process do not exist in the hydrogen plasma process. In the latter case, the hematite is melted and reduced simultaneously. Also, the net energy balance of iron oxide reduction via hydrogen plasma is exothermic and not endothermic as in the case of the solid-sate direct reduction. This fact ensures higher efficiency in power consumption of the former process.

3) The gangue elements of the hematite are gradually removed during reduction, due to their high vapor pressure. Silicon enrichments was identified in the interdendritic fayalite, at the wüstite/α-iron hetero-interfaces and in the primarily solidified oxide particles. Since silicon oxide reduction is not readily feasible, it is majorly evaporated in the form of SiO gas. Phosphorus and sulfur concentrations were also remarkably reduced from 5 min onwards and the final obtained iron is nearly free of gangue elements.

Our observations shed new light on the underlying thermodynamics and kinetics of hydrogen plasma-based reduction of iron ores, providing new insights into the feasibility of such approach as a carbon-neutral alternative for green iron production.

**Acknowledgements**

We thank Monika Nellessen and Katja Angenendt for their support to the metallography lab and SEM facilities at MPIE. We are grateful to Uwe Tezins, Christian Broß, and Andreas Sturm for their support to the FIB & APT facilities at MPIE. We are grateful to Benjamin Breitbach for the support to the X-ray diffraction facilities at MPIE. BG acknowledges financial support from the ERC-CoG-SHINE-771602.

**Figures**

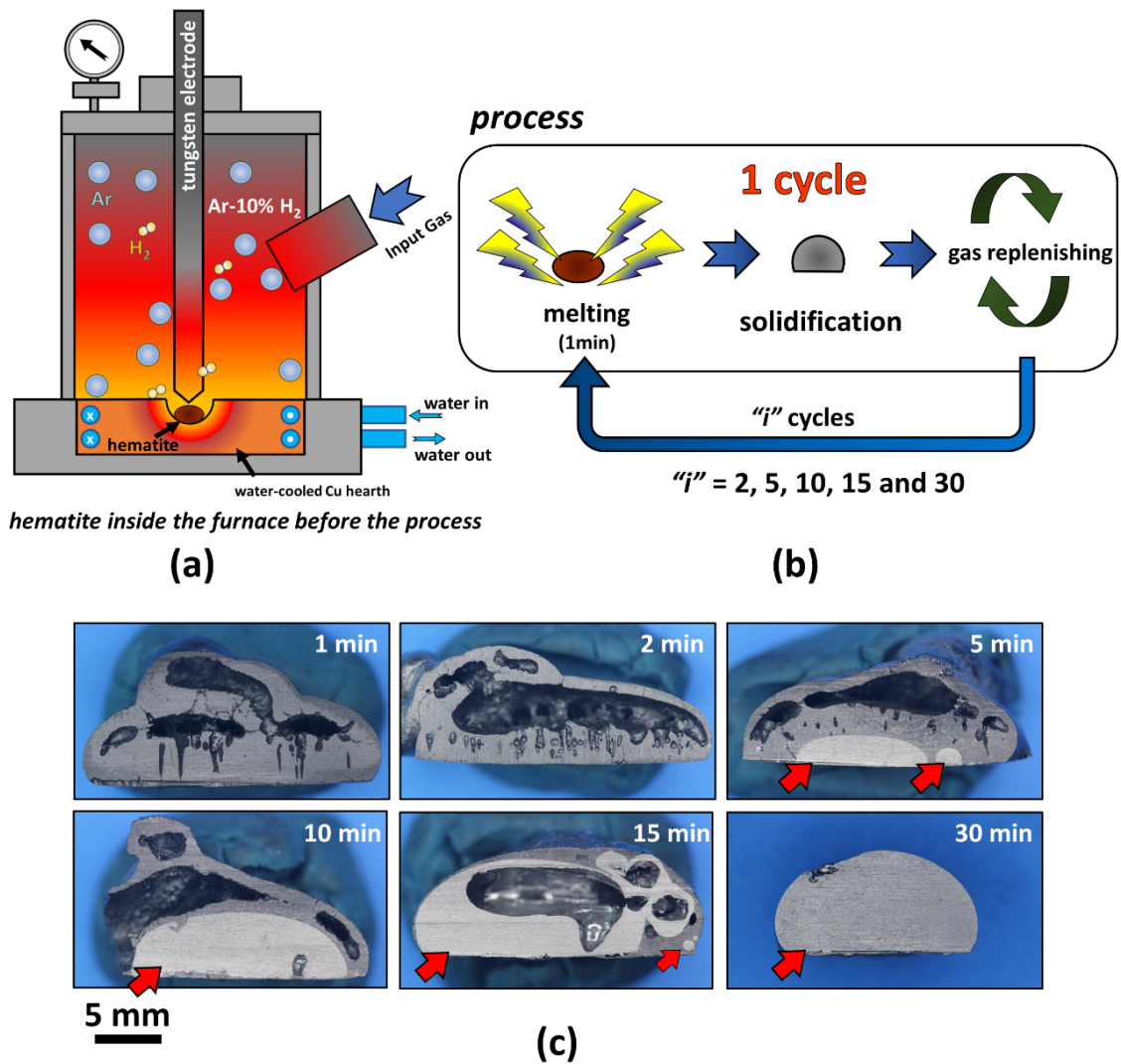

**Figure 1. (a)** Schematic illustration of the arc melting furnace equipped with a tungsten electrode and charged with Ar-10%$H_2$ gas mixture. The input hematite is placed on the water-cooled copper hearth inside the furnace before the reduction process. **(b)** Sequential steps for the reduction process. **(c)** Hematite pieces reduced with hydrogen plasma under different plasma exposure times: 1 min; 2 min; 5 min; 10 min; 15 min and 30 min. The red arrows indicate the presence of millimetric-scale transformed iron in the bottom of the samples. The upper part of the samples corresponds to the untransformed remaining oxide portion.

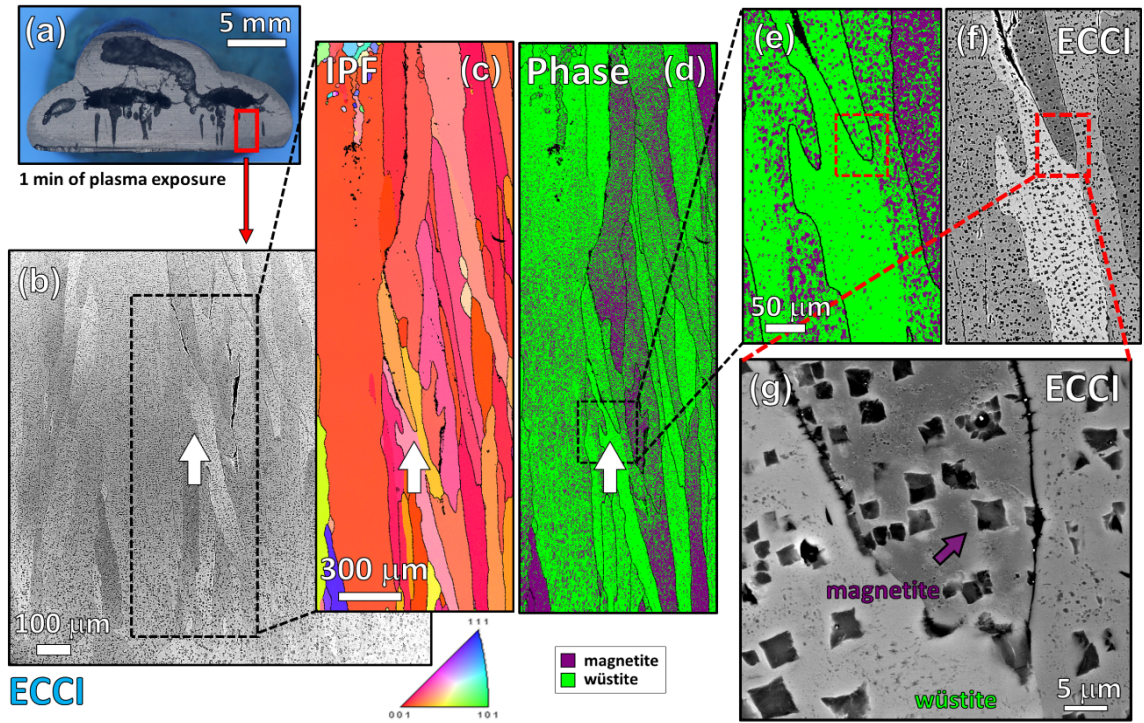

**Figure 2.** Microstructural characterization of the sample partially reduced for 1 min. **(a)** Overview of the sample. The red frame shows the chosen area for the correlative EBSD-ECCI probing approach. **(b)** ECCI-image of the area delimited by the red frame in (a). **(c)** EBSD inverse pole figure (IPF) of the region highlighted by the dashed frame in (b). The crystallographic orientations shown in this map are perpendicular to the normal direction of the sample, i.e. parallel to the solidification direction; **(d)** corresponding EBSD phase map showing magnetite and wüstite in purple and green, respectively. **(e)** Enlarged view of the area delimited by the frame in (d); **(f)** corresponding ECCI-image of the area displayed in (e). **(g)** Enlarged view of the area delimited by the red frame in (f), revealing the presence of cuboidal-shaped remaining magnetite dispersed in a columnar-shaped wüstite grain.

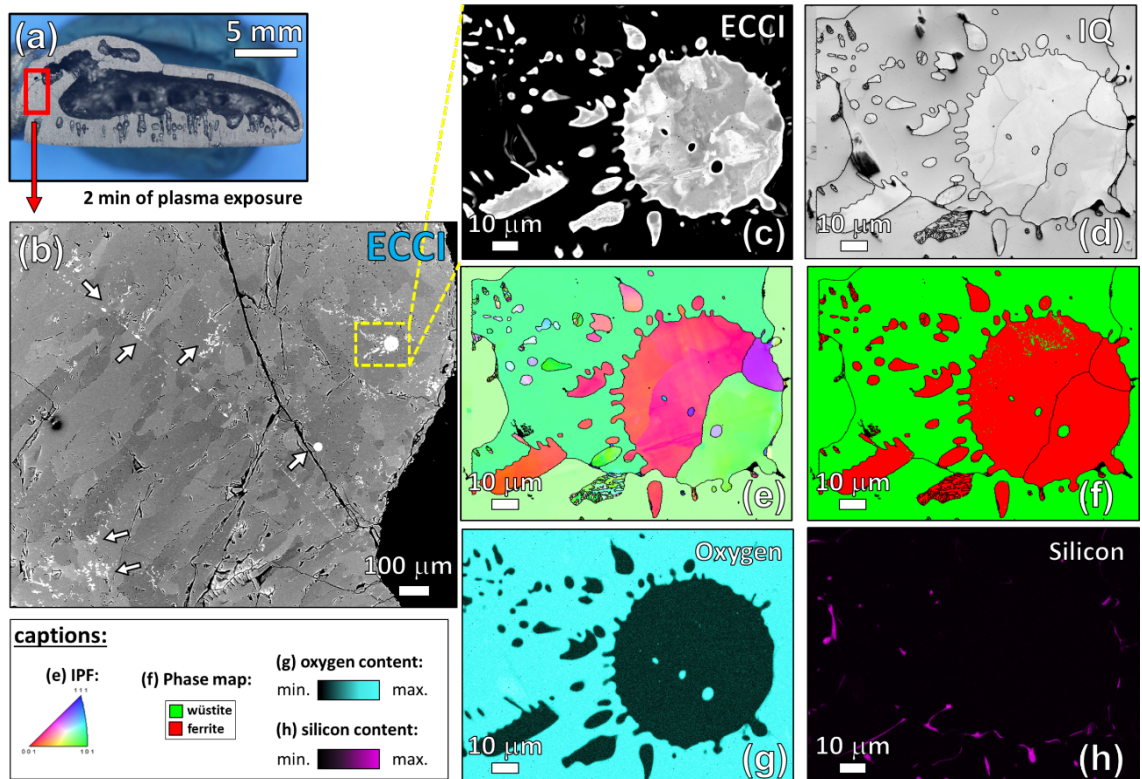

**Figure 3.** Microstructural characterization of the sample partially reduced for 2 min. **(a)** Overview of the sample. The red frame shows the chosen area for the correlative EBSD-ECCI probing approach. **(b)** ECCI-image of the area delimited by the red frame in (a). The white arrows evidence the presence of small domains of iron. **(c)** ECCI-image showing an enlarged view of the region delimitated by the yellow frame in (a). The corresponding EBSD maps of the area displayed in (c) are as follows: **(d)** image quality (IQ) map; **(e)** inverse pole figure (IPF) map in which the crystallographic orientations are shown perpendicular to the normal direction of the sample, i.e. parallel to the solidification direction; **(f)** phase map where wüstite and ferrite are represented by green and red respectively; **(g)** oxygen distribution and **(h)** silicon distribution maps. The corresponding captions are properly indicated in the bottom left-hand side of the figure.

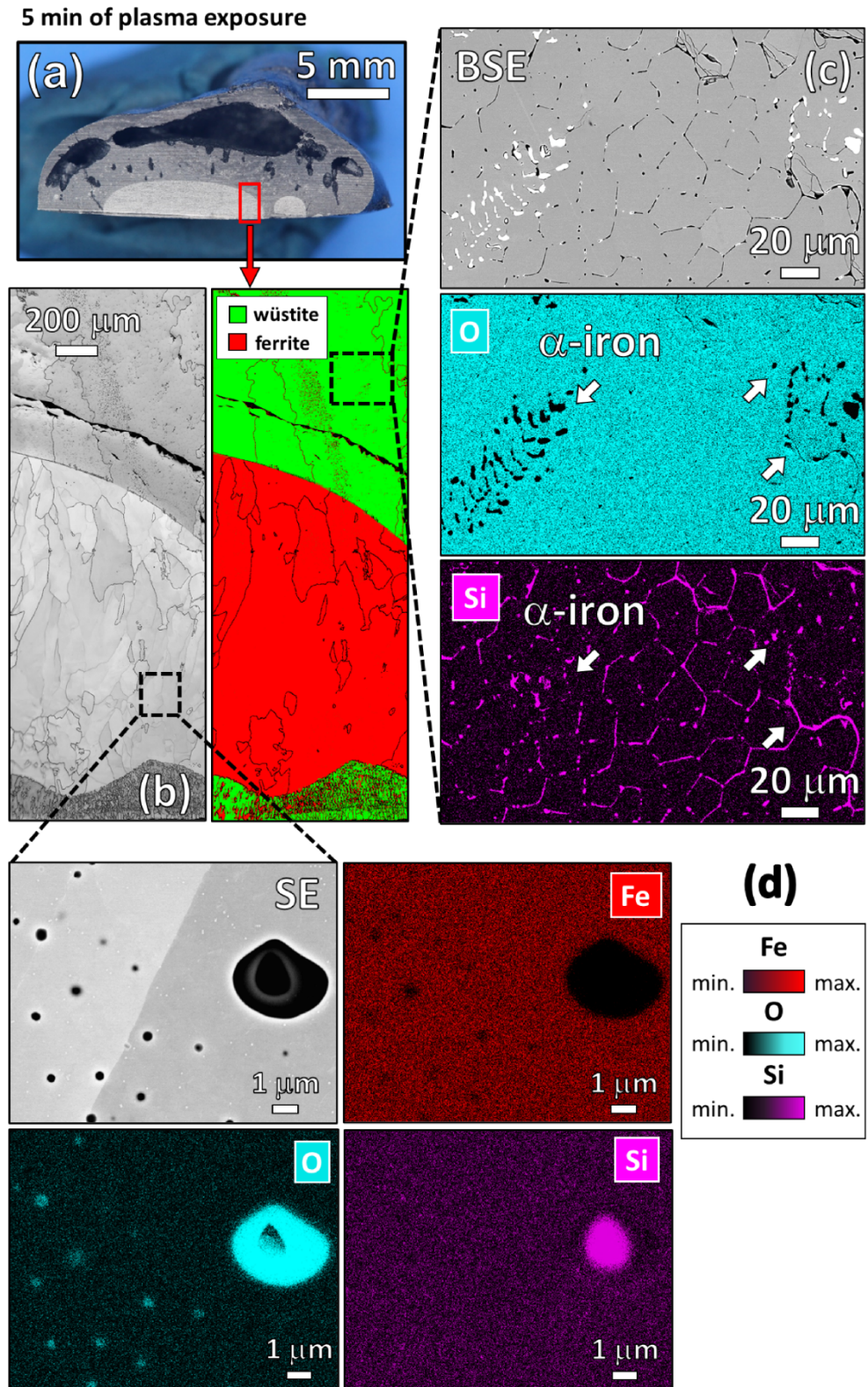

**Figure 4.** Microstructural characterization of the sample partially reduced for 5 min. **(a)** Overview of the sample. The red frame shows the area evaluated via EBSD whose maps are displayed in **(b)**. **(c)** Backscatter diffraction electrons (BSE) image of the frame displayed in (b). The oxygen and silicon distribution maps are also shown in (c). **(d)** Secondary electrons (SE) image of the interior of iron, revealing oxide particles. The corresponding iron, oxygen and silicon distribution maps are also shown in this image.

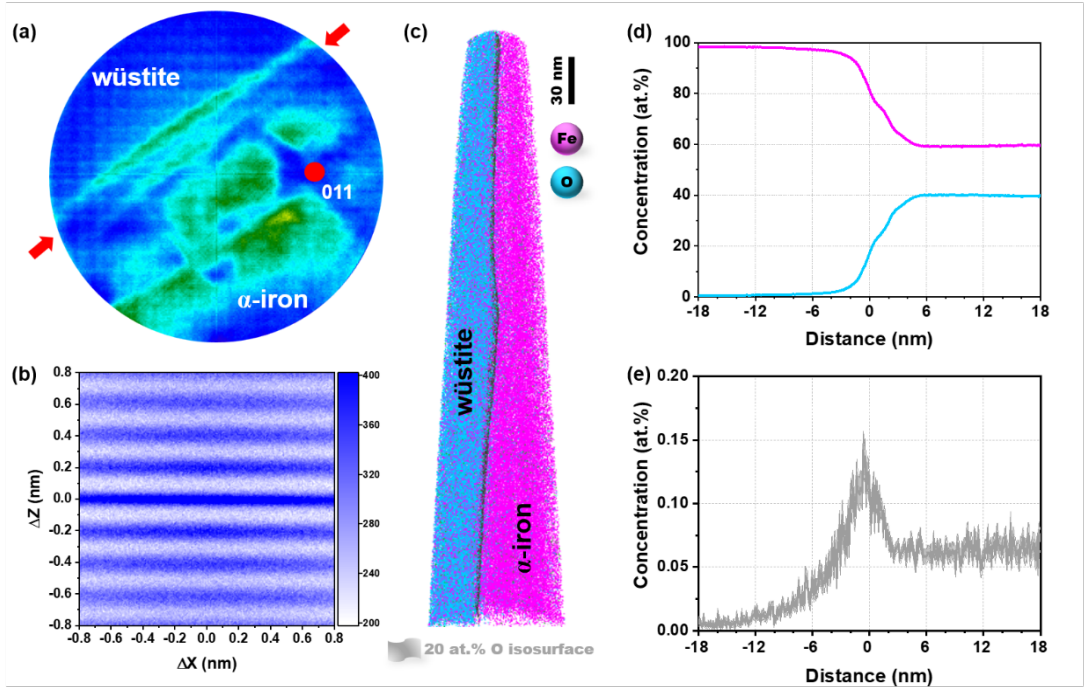

**Figure 5** Atom probe tomography (APT) analysis of the wüstite-iron interface in the partially reduced sample for 5 min. **(a)** Cumulative field evaporation histogram with the (011) pole of the reduced iron (the arrows indicate the position of the interface); **(b)** spatial distribution map analysis showing the {011} lattice plane of the reduced iron; **(c)** three-dimensional atom probe tomography maps of iron and oxygen, and the wüstite-iron interface is marked by a 20 at.% oxygen iso-concentration surface; concentration profiles of **(d)** iron and oxygen, as well as **(e)** silicon relative to the position of the wüstite-iron interface.

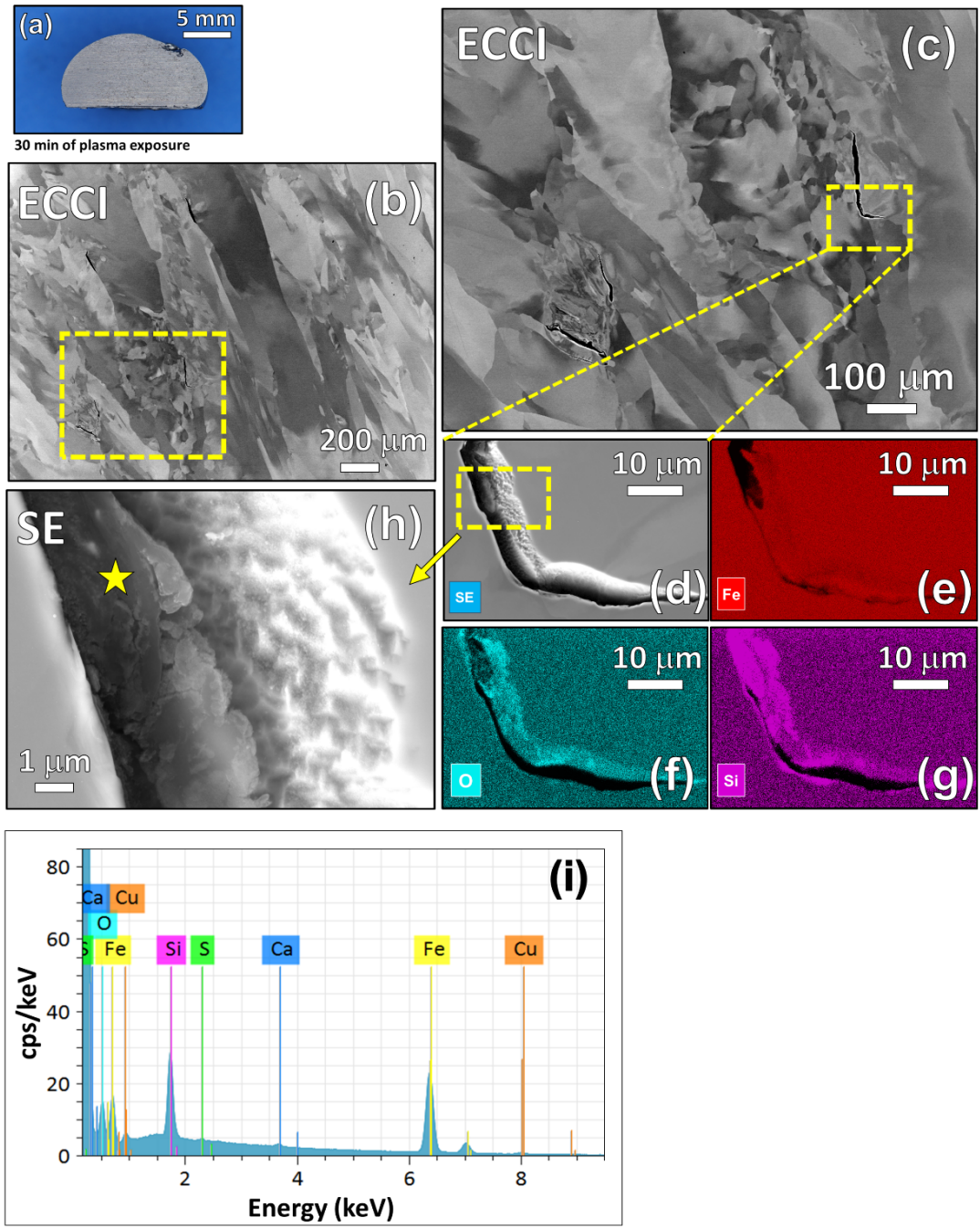

**Figure 6.** Microstructural characterization of the sample completely reduced after 30 min of exposure to hydrogen plasma. **(a)** Overview of the sample. **(b)** Corresponding microstructure observed with the aid of the ECCI technique. **(c)** Enlarged view of the region delimitated by the yellow frame in (b). **(d)** Enlarged view of the area delimited by the yellow frame displayed in (c), showing details of a worm-like inclusion. Corresponding elemental maps for **(e)** iron; **(f)** oxygen and **(g)** silicon. **(h)** Morphological details of the inclusion displayed in (d). The yellow star represents the area probed via EDS. The obtained elemental spectrum is shown in **(i)**.

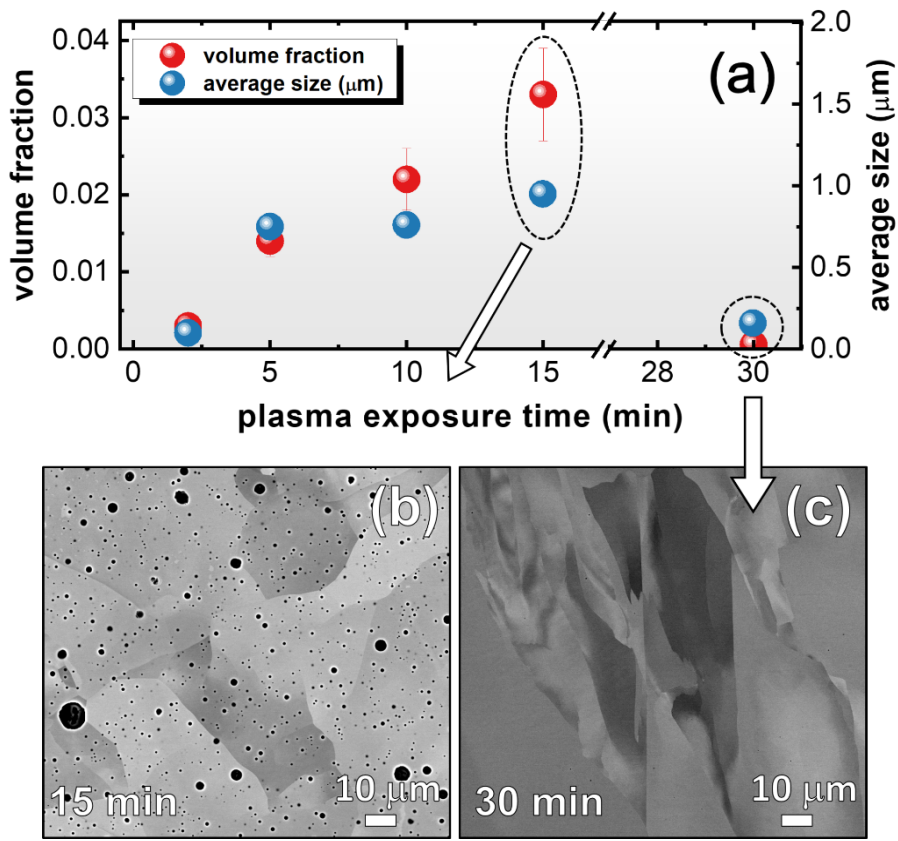

**Figure 7. (a)** Evolution of the volume fraction and average size of the oxide particles pinned inside iron. **(b)** ECCI-image of the microstructure of the iron formed in the 15-min partially reduced sample **(c)** ECCI-image of the sample completely reduced after 30 min of exposure to the hydrogen plasma.

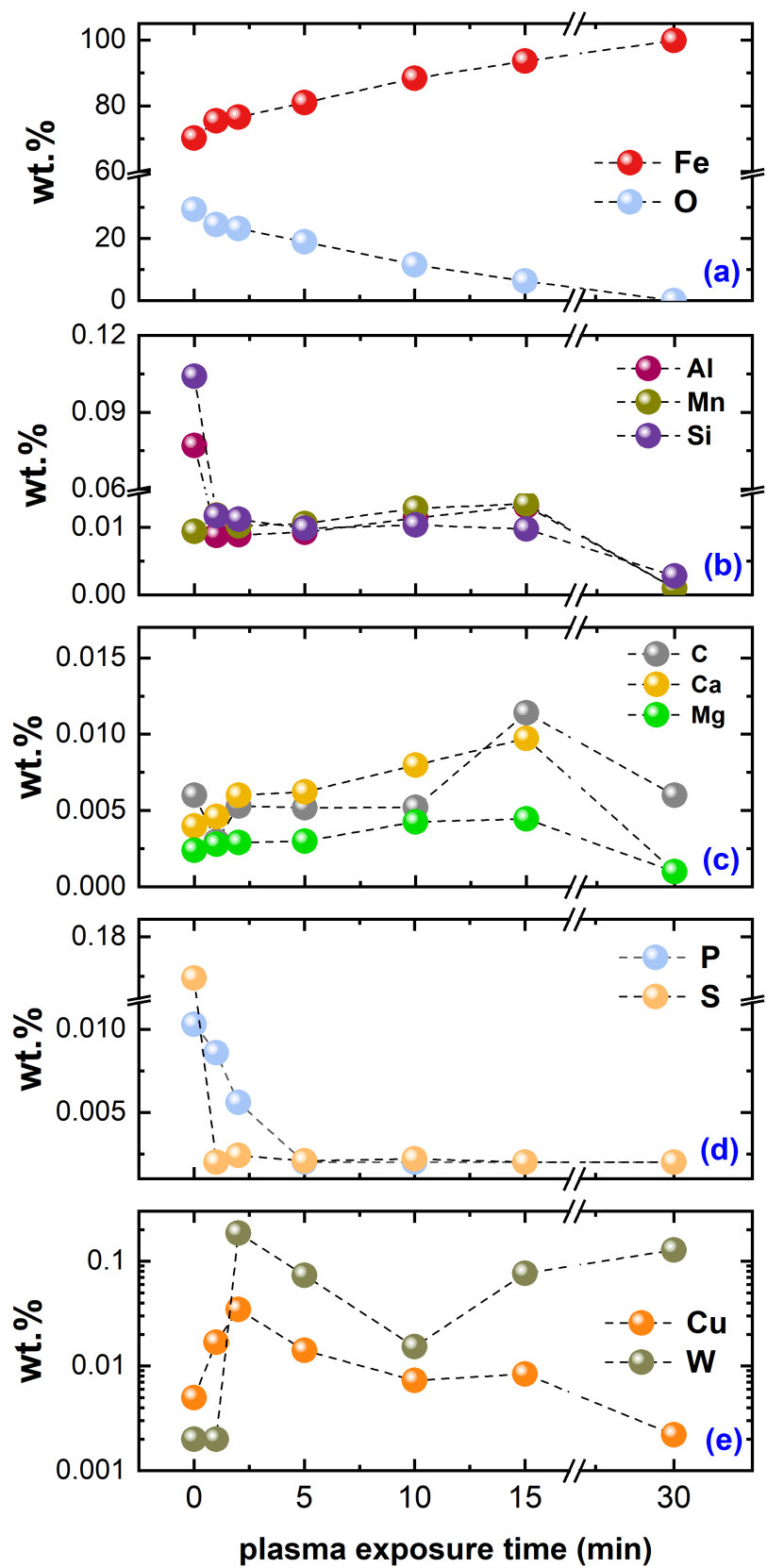

**Figure 8.** Chemical composition evolution during reduction. Changes in weight percentage in terms of **(a)** Fe and O; **(b)** Al, Mn, and Si; **(c)** C, Ca, and Mg; **(d)** P and S; **(e)** Cu and W, where the logarithmic y-axis serves to facilitate the visualization.

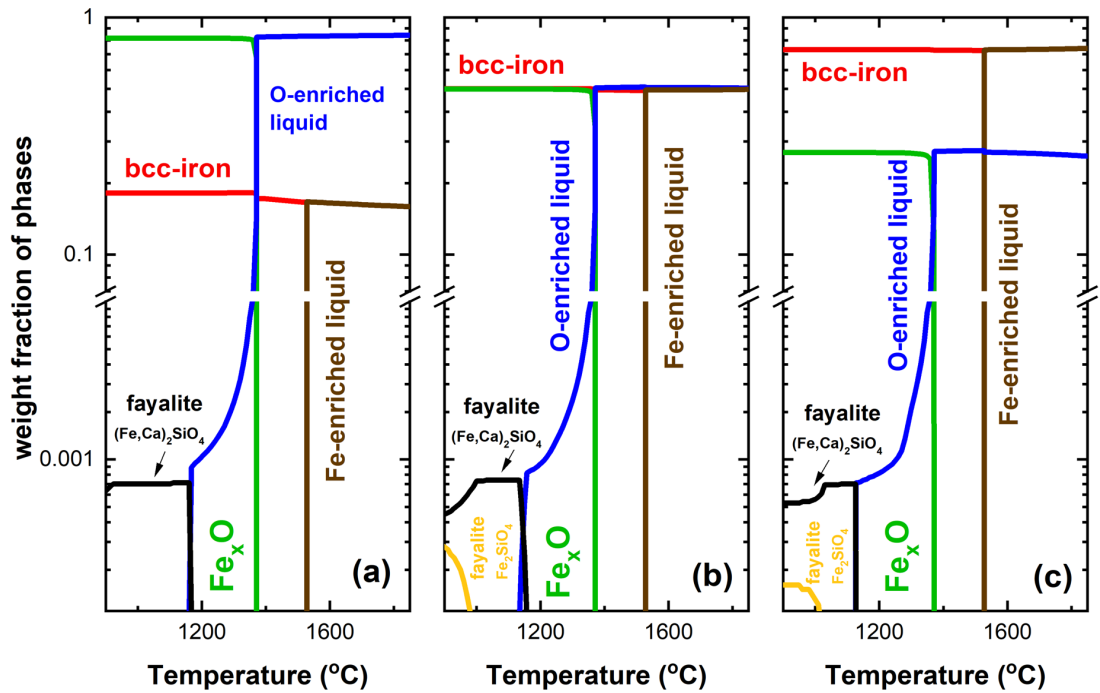

**Figure 9.** Metastability calculations for the samples partially reduced for **(a)** 5 min, **(b)** 10 min and **(c)** 15 min, conducted using the software ThermoCalc with the aid of the metal oxide solutions database TCOX10. For the construction of these diagrams, austenite (fcc-iron) was excluded from the calculations.

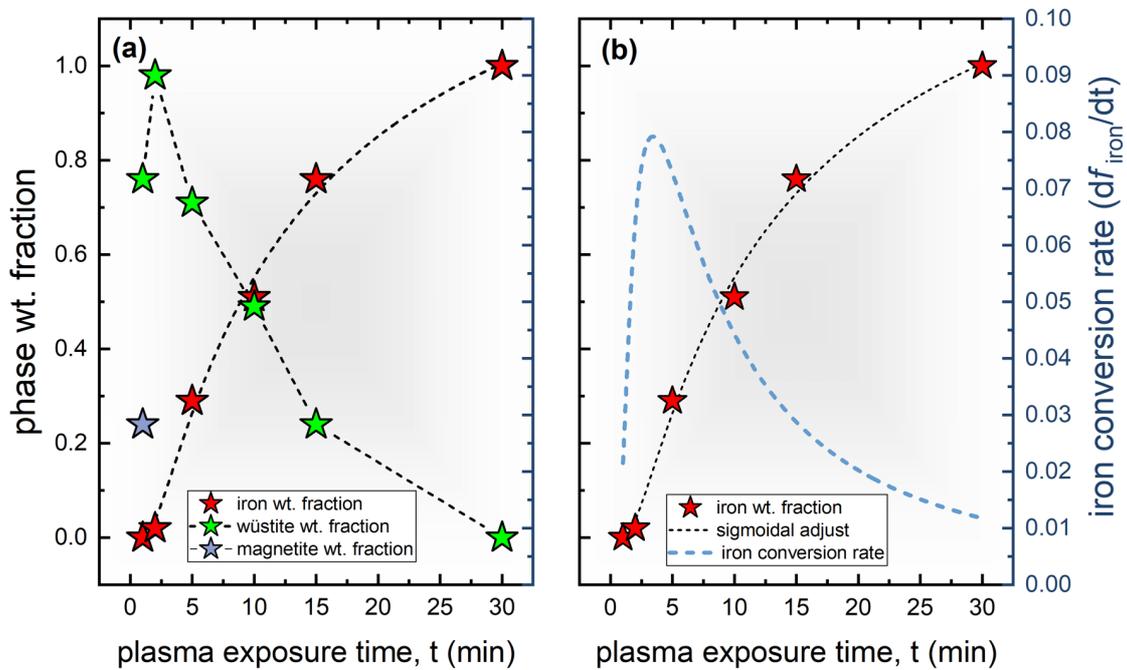

**Figure 10.** Phase transformation kinetics during hydrogen plasma-based reduction of the 9 g-hematite pieces. **(a)** Changes in the phase weight fractions of magnetite, wüstite, and ferrite. **(b)** Iron conversion kinetics and corresponding transformation rates.

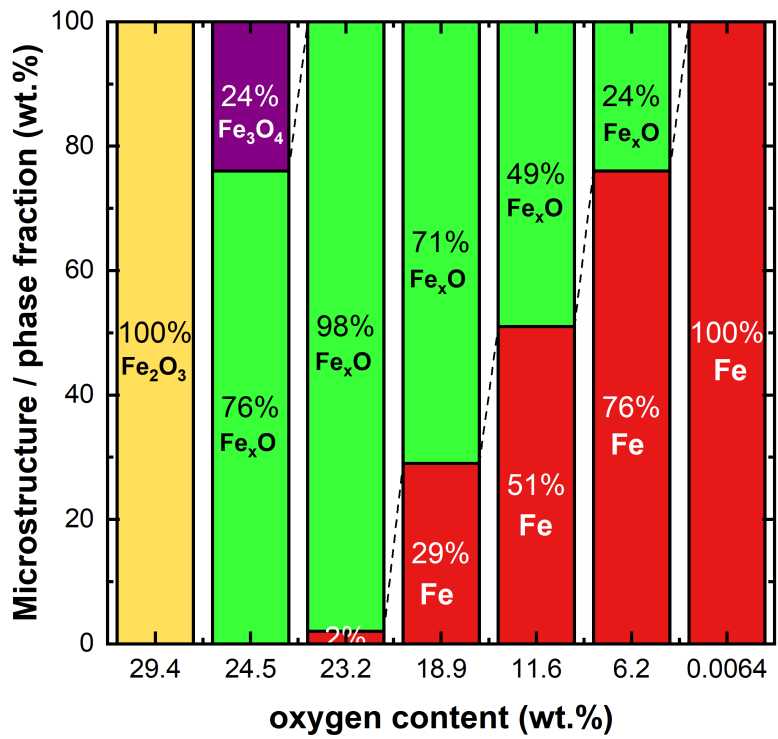

**Figure 11.** Diagram 'microstructure versus oxygen content' for the 9 g-hematite pieces reduced via hydrogen plasma under different exposure times.

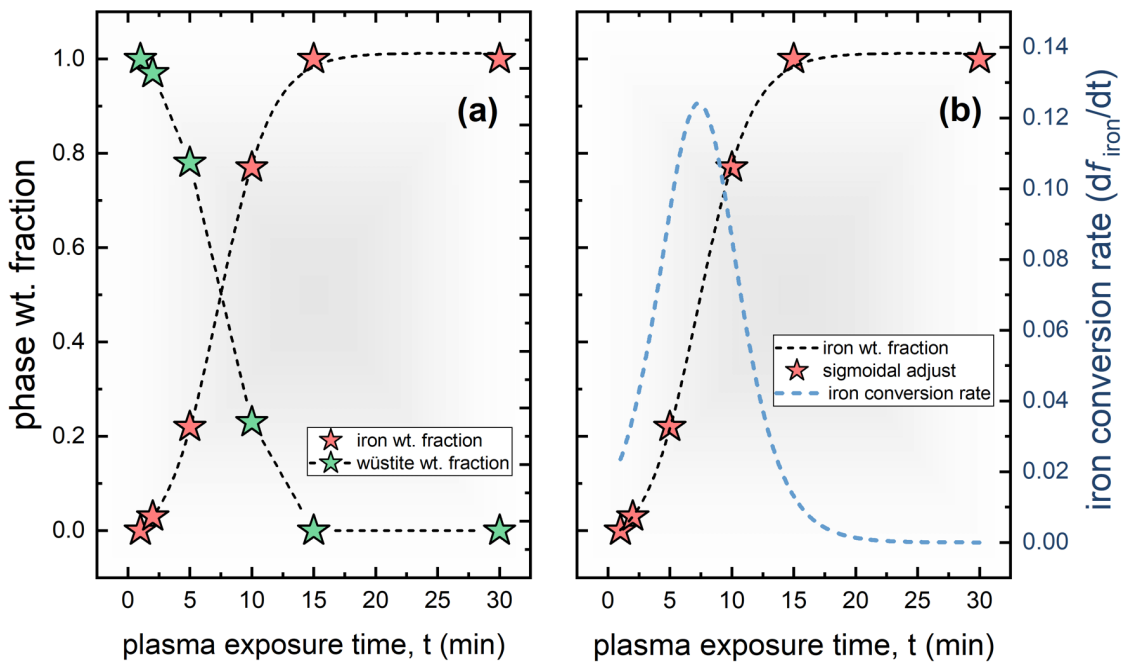

**Figure 12.** Phase transformation kinetics during hydrogen plasma-based reduction of the 15 g-hematite pieces. **(a)** Changes in the phase weight fractions of magnetite, wüstite, and ferrite **(b)** Iron conversion kinetics and corresponding transformation rates.

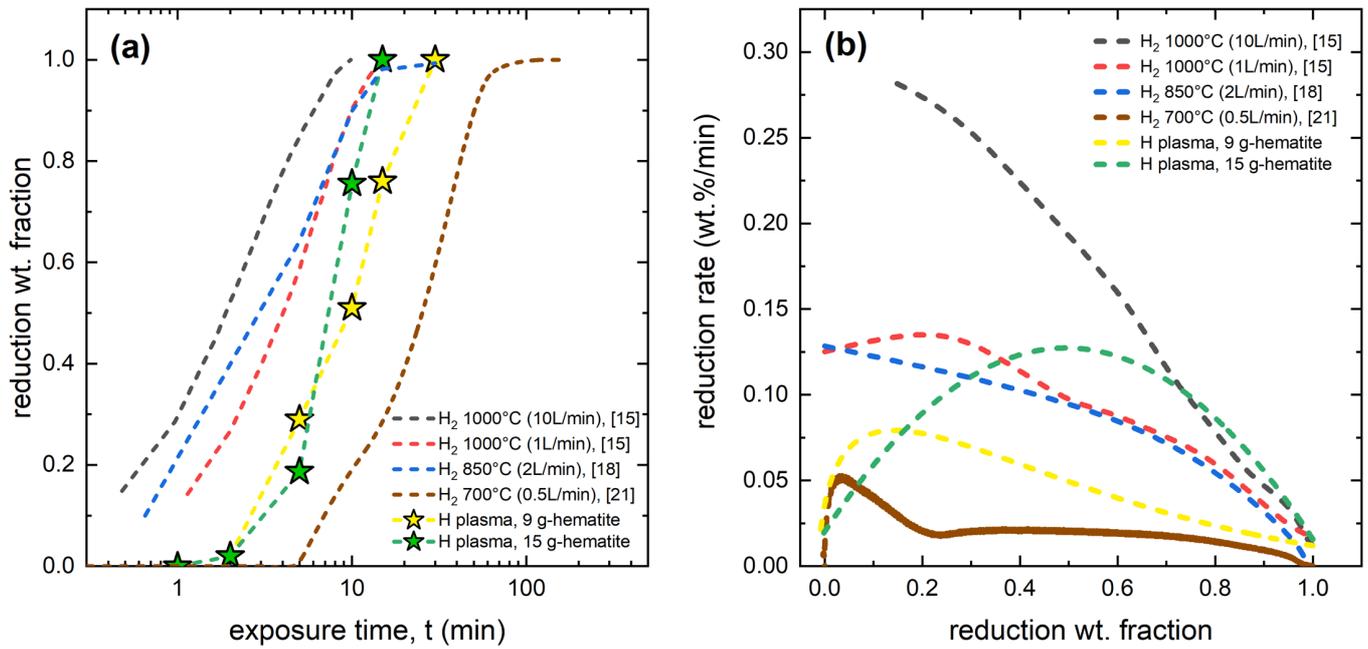

**Figure 13. (a)** Iron conversion kinetics during hydrogen plasma-based reduction of the 9 and 15 g-hematite pieces conducted in this study, represented by the yellow and green curves, respectively. These plasma synthesis results are compared to the direct reduction kinetics of a hematite pellet via solid-state direct reduction using molecular hydrogen at 700°C (brown curve) reported in a preceding study [21]; direct reduction kinetics using molecular hydrogen at 1000°C with $H_2$ flow rates of 10 L/min (black line) and 1 L/min (red line) from Ref. [15], and the direct reduction kinetics of iron ore pellets with molecular hydrogen at 850°C (2 L/min) from Ref. [18]. **(b)** Comparison of the corresponding reduction rates vs. the reduction wt. fractions.

**Table content**

**Table 1.** Chemical composition of the initial hematite measured by inductively coupled plasma optical emission and infrared absorption spectroscopies (in wt.%).

| Fe | Si | Al | Na | Mn | Cu | Ca | Mg | Ti |
|---|---|---|---|---|---|---|---|---|
| 70.228 | 0.104 | 0.077 | 0.017 | 0.0094 | 0.005 | 0.004 | 0.0024 | 0.0019 |
| W | C | P | S | O | | | | |
| 0.002 | 0.006 | 0.0103 | 0.133 | 29.4 | | | | |

**Table 2.** Chemical composition of reduced bcc-iron and wüstite of the sample reduced for 5 min analyzed by atom probe tomography (in at.%).

| | Fe | Al | Ca | Mg | Mn | O | Si | Co | Ni | V |
|---|---|---|---|---|---|---|---|---|---|---|
| **bcc-iron** | 98.082 ±0.302 | 0.006 ±0.002 | 0.0003 ±0.0006 | 0.004 ±0.002 | 0.035 ±0.006 | 0.742 ±0.175 | 0.013 ±0.009 | 0.005 ±0.002 | 0.003 ±0.002 | 0.004 ±0.002 |

|  | Fe | Al | C | Ca | Cu | Mg | Mn | O | P | S | Si | W |
|---|---|---|---|---|---|---|---|---|---|---|---|---|
| wüstite | 59.351 ±0.219 | 0.083 ±0.009 | 0.0016 ±0.0011 | 0.044 ±0.006 | 0.060 ±0.007 | 39.893 ±0.206 | 0.063 ±0.007 | 0.035 ±0.006 | 0.007 ±0.002 | 0.071 ±0.007 | | |

**Table 3.** Chemical composition evolution of the produced iron measured by inductively coupled plasma optical emission and infrared absorption spectroscopies (in wt.%).

|  | Fe | Al | C | Ca | Cu | Mg | Mn | O | P | S | Si | W |
|---|---|---|---|---|---|---|---|---|---|---|---|---|
| **5 min** | bal. | 0.008 | 0.005 | 0.004 | 0.019 | 0.003 | 0.009 | 9.590 | <0.002 | <0.002 | 0.009 | 0.055 |
| **10 min** | bal. | <0.001 | 0.005 | <0.001 | 0.014 | <0.001 | <0.001 | 0.500 | <0.002 | <0.002 | 0.002 | 0.002 |
| **15 min** | bal. | <0.001 | 0.005 | <0.001 | 0.011 | <0.001 | <0.001 | 0.500 | <0.002 | <0.002 | 0.002 | 0.041 |
| **30 min** | bal. | <0.001 | 0.006 | <0.001 | 0.002 | <0.001 | <0.001 | 0.006 | <0.002 | <0.002 | 0.003 | 0.128 |

**Table 4.** Oxygen removal during reduction and chemical composition of the furnace atmosphere. Total amount of hydrogen provided in the furnace chamber prior to reduction. Excess of hydrogen after a certain exposure time to the plasma. Corresponding fraction of $H_2$ in the mixture of $H_2 + H_2O$ after reduction.

| Exposure time (min) | Oxygen removal (%) | Total provided $H_2$ (mol) | Excess $H_2$ (mol) | $H_2$ in ($H_2 + H_2O$) |
|---|---|---|---|---|
| 1 | 16.7 | 0.065 | 0.041 | 0.63 |
| 2 | 21.1 | 0.131 | 0.099 | 0.76 |
| 5 | 35.7 | 0.327 | 0.271 | 0.83 |
| 10 | 60.5 | 0.654 | 0.551 | 0.84 |
| 15 | 78.9 | 0.981 | 0.836 | 0.85 |
| 30 | ~100 | 1.962 | 1.558 | 0.79 |

**Supplementary Material**

# Sustainable steel through hydrogen plasma reduction of iron ore: process, kinetics, microstructure, chemistry

I. R. Souza Filho[1,*], Y. Ma[1], M. Kulse[1], D. Ponge[1], B. Gault[1,2], H. Springer[1,3], D. Raabe[1]

[1] *Max-Planck-Institut für Eisenforschung, Max-Planck-Str. 1, 40237 Düsseldorf, Germany*
[2] *Department of Materials, Imperial College London, South Kensington, London SW7 2AZ, UK*
[3] *Institut für Bildsame Formgebung, RWTH Aachen University, Intzestr. 10, 52072 Aachen, Germany*

*corresponding author: i.souza@mpie.de

**1. Thermodynamic simulations (phase equilibrium)**

Figure S.1 shows the phase diagrams for the samples partially reduced for 5, 10, and 15 min. These diagrams were calculated taking into account the chemical composition of the samples (Fig. 8 of the manuscript) using the software ThermoCalc coupled with the database TCOX10. Differently from Fig. 9 of the manuscript, the face-centered cubic iron (fcc-iron) was included in the calculations of the diagrams shown in Fig. S.1.

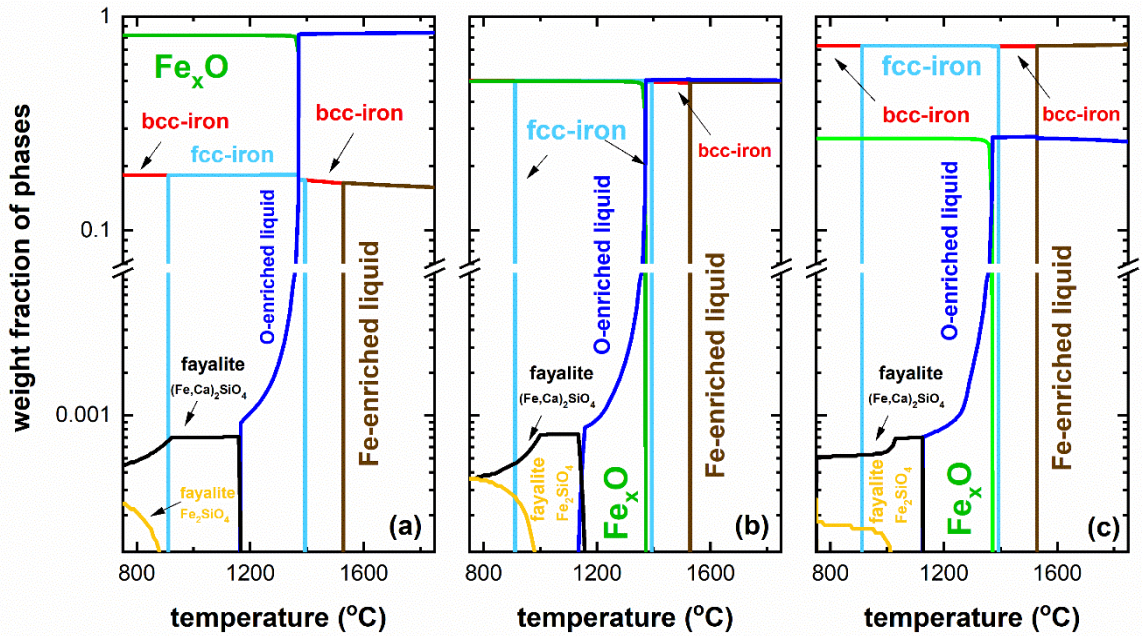

**Figure S.1.** Phase equilibrium calculations for the samples partially reduced for **(a)** 5 min, **(b)** 10 min and **(c)** 15 min, conducted using the software ThermoCalc with the aid of the metal oxide solutions database TCOX10.

## 2. Site-specific lift out of APT specimens

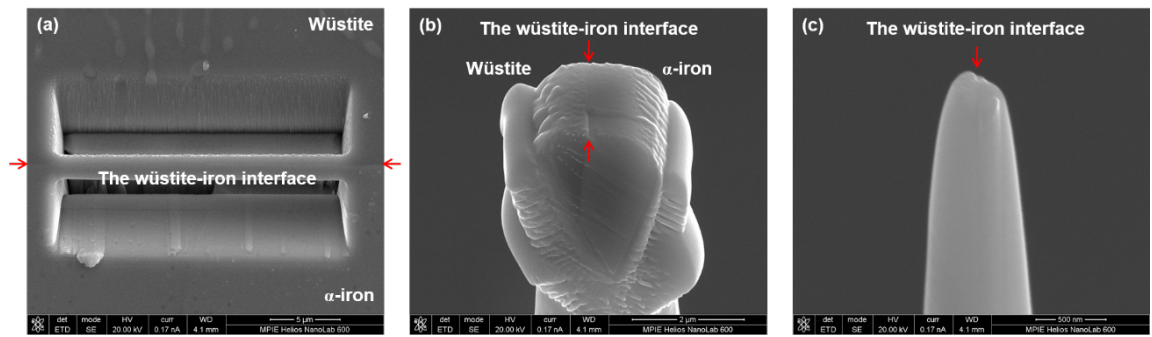

**Figure S.2.** Precision site-specific sample preparation for atom probe tomography tips: **(a)** wedge milled along the wüstite-iron interface of the 5 min-reduced sample; **(b)** lift-out sample on the pre-fabricated Si micro-post for annular milling; **(c)** APT tip before the final annular milling step.

## 3. Porosity

Figure S.3 shows the porous area percentage estimated via quantitative metallography for the partially reduced samples with exposure times to the plasma varying from 1 to 15 min.

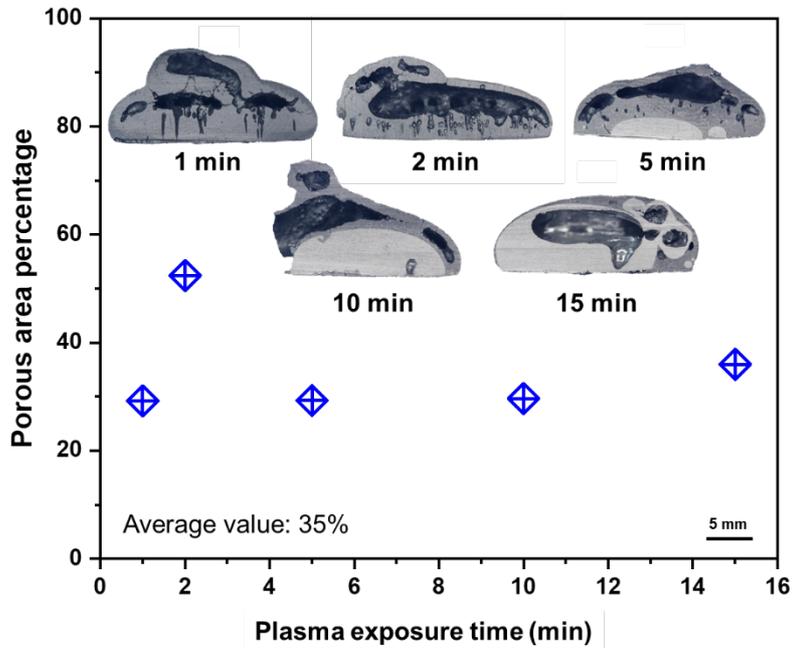

**Figure S.3.** Porous area percentage of the samples partially reduced under hydrogen plasma.

## 4. Sample partially reduced for 10 min
### 4.1 Microstructural characterization

Two distinct regions of the sample partially reduced for 10 min were analyzed. Fig. S.4 (a) shows a first area close to the iron/remaining oxide interface (red square). Figs. S.4 (b) and (c) show the corresponding IQ and phase maps acquired from the region marked by the red frame in Fig. S.4 (a). This figure shows that the metallic iron drops (arrowed) are found within the remaining wüstite. An enlarged view of the phase interface indicated by the blue frame in Fig. S.4 (c) is given by the secondary electron (SE) image in Fig. S.4 (d) and its corresponding EDS elemental maps, which show the local iron (Fig. S.4 e), oxygen (Fig. S.4 f) and silicon (Fig. S.4 g) distributions. In Figs. S.4 (d-g), oxides particles within iron are indicated by arrows and the Si-enriched interdendritic structures are observed in wüstite. Analyzes performed via XRD measurements (Fig. S.6) reveal that this minor interdendritic constituent is fayalite ($Fe_2SiO_4$).

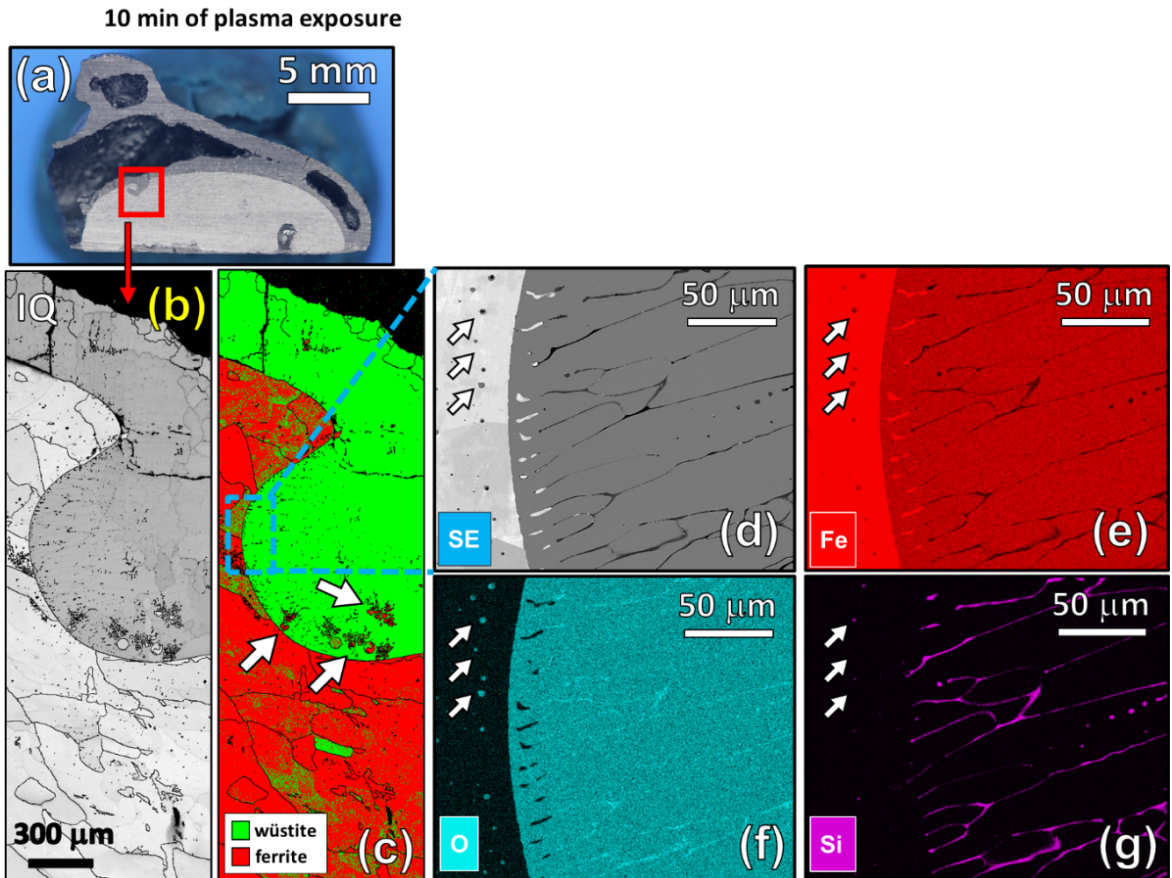

**Figure S.4.** Microstructural characterization of the sample partially reduced for 10 min. **(a)** Overview of the sample. The red frame shows the chosen area for the correlative EBSD-EDS probing approach. **(b)** EBSD image quality (IQ) map for the area delimited by the red frame in S.3 (a); **(c)** corresponding phase map where wüstite and ferrite are represented by green and red. respectively. **(d)** secondary electron (SE) image of the region highlighted by the blue frame in S.4 (c). Corresponding EDS maps showing the local distribution of iron **(e)**, oxygen **(f)** and silicon **(g)**.

A second area within the iron is shown in Fig. S.5 (a) (yellow square). The large iron portion has elongated grains inclined in approximately 45º related to solidification direction, as shown by the IQ and IPF maps (Fig. S.5 b and c) as well as by the ECCI-image displayed in Fig. S.5 (d). A higher magnification of this figure in combination with EDS mapping enables a qualitative evaluation of the chemical nature of the dispersed oxide particles within iron. Some of them present a Si-enriched core and a Fe- and O-bearing shell, while others seem to be fully composed of either Si and O or Fe and O.

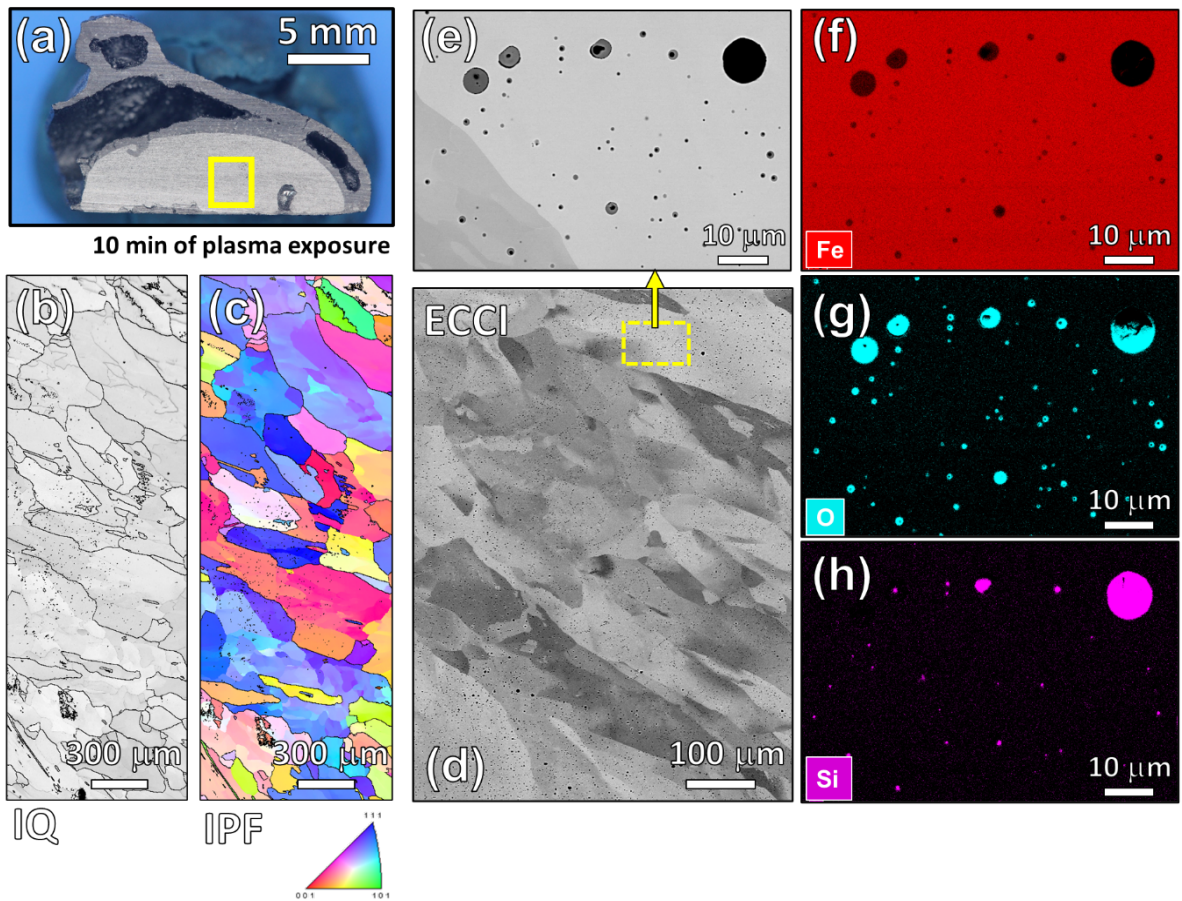

**Figure S.5.** Microstructural characterization of the sample partially reduced for 10 min. **(a)** Overview of the sample. The yellow frame highlights the region where the correlative EBSD. ECCI and EDS analysis was carried out. **(b)** EBSD image quality (IQ) map for the area delimited by the red frame in S.4 (a); **(c)** corresponding inverse pole figure (IPF). The crystallographic directions are shown parallel to the solidification direction. **(d)** ECCI image of the iron portion. **(e)** Enlarged view of the iron region evidencing the presence of particles. Corresponding EDS elemental maps shows the local iron **(f)**. oxygen **(g)** and silicon **(h)** distributions. respectively.

## 4.2 X-ray measurements

Figure S.6 shows a representative diffractogram of the oxide powder obtained from the sample partially reduced for 10 min. In this figure, the black line represents the experimentally determined diffractogram. The red, blue, and green lines represent the deconvoluted peaks of wüstite, ferrite and fayalite, respectively. Rietveld refinement reveals a weight fraction of 0.93, 0.04 and 0.03 for these constituents, respectively.

The total mass of the oxide portion for these sample is 0.978 g. This means that the corresponding mass of fayalite in the sample is 0.029 g. Therefore, the weight percentage of fayalite when considering the entire sample (4.036 g, Fig. 1c of the manuscript) is 0.73%.

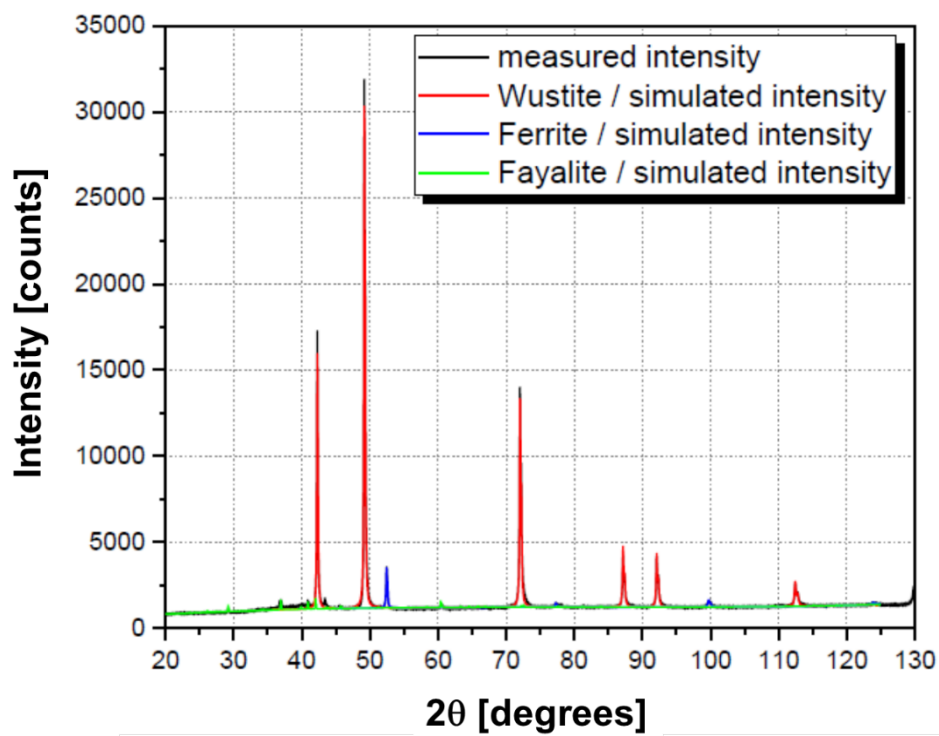

**Figure S.6.** Representative diffractogram of the sample partially reduced for 10 min. The black line represents the experimentally determined data. The red, blue, and green lines represent wüstite, ferrite and fayalite deconvoluted peaks, obtained after Rietveld refinement.